\documentclass[twocolumn,tighten]{aastex63}
\usepackage{graphicx}
\usepackage{amsmath}
\usepackage{natbib}
\usepackage{amssymb}

\usepackage{sidecap}

\shorttitle{Luminous FBOTs and SNe Ibn/Icn from WR/BH Mergers}
\shortauthors{B.~D.~Metzger}

\begin{document}

\newcommand{\be}{\begin{equation}}
\newcommand{\ee}{\end{equation}}

\title{Luminous Fast Blue Optical Transients and Type Ibn/Icn SNe from Wolf-Rayet/Black Hole Mergers}

\author[0000-0002-4670-7509]{Brian D. Metzger}
\affil{Department of Physics and Columbia Astrophysics Laboratory, Columbia University, Pupin Hall, New York, NY 10027, USA}
\affil{Center for Computational Astrophysics, Flatiron Institute, 162 5th Ave, New York, NY 10010, USA}

\begin{abstract}

Progenitor models for the ``luminous'' subclass of Fast Blue Optical Transients (LFBOTs; prototype: AT2018cow) are challenged to simultaneously explain all of their observed properties: fast optical rise times $\lesssim$ days; peak luminosities $\gtrsim 10^{44}$ erg s$^{-1}$; low yields $\lesssim 0.1M_{\odot}$ of $^{56}$Ni; aspherical ejecta with a wide velocity range ($\lesssim 3000$ km s$^{-1}$ to $\gtrsim 0.1-0.5 c$ with increasing polar latitude); presence of hydrogen-depleted-but-not-free dense circumstellar material (CSM) on radial scales from $\sim 10^{14}$ cm to $\sim 3\times 10^{16}$ cm; embedded variable source of non-thermal X-ray/$\gamma-$rays, suggestive of a compact object.  We show that all of these properties are consistent with the tidal disruption and hyper-accretion of a Wolf-Rayet (WR) star by a black hole (BH) or neutron star (NS) binary companion.  In contrast with related previous models, the merger occurs with a long delay ($\gtrsim 100$ yr) following the common envelope (CE) event responsible for birthing the binary, as a result of gradual angular momentum loss to a relic circumbinary disk.  Disk-wind outflows from the merger-generated accretion flow generate the $^{56}$Ni-poor aspherical ejecta with the requisite velocity range.  The optical light curve is powered primarily by reprocessing X-rays from the inner accretion flow/jet, though CSM shock interaction also contributes.  Primary CSM sources include WR mass-loss from the earliest stages of the merger ($\lesssim 10^{14}$ cm) and the relic CE disk and its photoevaporation-driven wind ($\gtrsim 10^{16}$ cm).  Longer delayed mergers may instead give rise to supernovae Type Ibn/Icn (depending on the WR evolutionary state), connecting these transient classes with LFBOTs.
\end{abstract}


\section{Introduction}

``Fast Blue Optical Transients" (FBOTs; \citealt{Drout+14,Arcavi+16,Pursiainen+18,Ho+21b}) are an emerging class of supernova(SN)-like stellar explosions characterized by optical rise times of only days and peak luminosities up to $\gtrsim 10^{44}$ erg s$^{-1}$, which in some cases approach or exceed those of superluminous supernovae (SLSNe; e.g., \citealt{Inserra19}).  This paper focuses on the most luminous subclass of FBOTs, hereafter denoted ``LFBOTs'', which are extremely rare (local volumetric rate $\sim 2-400$ Gpc$^{-3}$ yr$^{-1}$, or $\lesssim 0.6\%$ of the core-collapse SN rate; e.g., \citealt{Coppejans+20,Ho+21b}).  The broader population of fast-evolving transients with lower luminosities are more common ($\sim 10^{4}$ Gpc$^{-3}$ yr$^{-1}$; e.g., \citealt{Drout+14,Ho+21b}) and likely possess a distinct origin from LFBOTs, instead representing extreme members along the continuum of other established SN types (e.g., \citealt{Ho+21b}).

\subsection{Lessons from AT2018cow and its Analogs}

The prototypical member of the LFBOT class is AT2018cow, which exhibited multi-wavelength emission spanning from radio to gamma-rays and occurred at a distance of only 60 Mpc \citep{Prentice+18,RiveraSandoval+18,Kuin+19,Margutti+19,Perley+19,Ho+19,Nayana&Chandra21}.  The optical emission from AT2018cow rose over just a few days to a peak luminosity $L_{\rm opt} \approx 4\times 10^{44}$ erg s$^{-1}$ before declining thereafter, roughly as $L_{\rm opt} \propto t^{-2}$.  The initial spectra at $t \lesssim 15$ days were mostly featureless and indicated large photosphere expansion velocities $v \gtrsim 0.1 c$ and temperatures $\sim 3\times 10^{5}$ K.  Later spectra revealed a persistent optically-thick photosphere and the emergence of H and He emission features, which abruptly decrease to velocities $v \sim 3000-4000$ km s$^{-1}$ with no evidence for ejecta cooling (e.g., \citealt{Perley+19,Margutti+19,Xiang+21}), quite unlike predictions for standard shock break-out emission (e.g., \citealt{Nakar&Sari10}).  However, the presence of narrow He emission lines $\lesssim 300$ km s$^{-1}$ exhibit similarities with SNe Ibn or transitional SNe Ibn/IIn objects and point to shock interaction between the ejecta and H-depleted circumstellar material (CSM; e.g., \citealt{Fox&Smith19,Dessart+21}).

AT2018cow also produced soft $\sim$ keV X-ray emission even from the early days of the explosion (e.g., \citealt{RiveraSandoval+18,Kuin+19}), which decayed gradually initially but later around day 20$-$the same time as the abrupt decrease in optical spectral velocities$-$began to track the steeper decay of the optical emission, with a comparable luminosity.  The X-ray light curve was highly variable and became increasingly so with time, decaying faster than $L_X \propto t^{-4}$ after about one month.   Hard  $\gtrsim 10$ keV X-ray emission was also detected \citep{Margutti+19} with a distinct spectrum characterized by broadened Fe emission lines and a spectral peak around 30 keV (similar to the ``Compton hump" feature observed in AGN accretion disk reflection spectra; e.g., \citealt{Reynolds99}) and which faded over the first weeks.  A quasi-periodic feature in the X-ray timing properties was recently reported at 225 Hz \citep{Pasham+21}.

As described by \citet{Margutti+19}, the many unusual properties of AT2018cow can be understood if the bulk of the emission is powered by a compact, highly time-variable X-ray source embedded within a highly aspherical ejecta shell.  The latter is comprised of a dense low-velocity $v_{\rm slow} \sim 3000$ km s$^{-1}$ equatorially-concentrated belt and a much faster $v_{\rm fast} \gtrsim 0.1 c$ polar-directed outflow of mass $M_{\rm fast} \sim 0.1 M_{\odot}$,  which controls the initial rise time of the transient following the usual \citet{Arnett82} relationship.  A similarly low polar ejecta mass is required for the central X-ray source to photoionize the polar material and enable the soft X-rays to escape at such an early phase.

The bulk of the optical emission is likely powered by partial reprocessing of the central X-ray source by the fast polar outflow (e.g., \citealt{Margutti+19,Piro&Lu20,Calderon+21,Chen&Shen22}), which carries out the photosphere at early times, giving rise to the broad lines (though the kinetic energy dissipated by shock interaction between the ejecta and dense CSM could also play a role).   Passive cooling of an earlier shock-heated envelope (e.g., \citealt{Gottlieb+22}) cannot readily explain the persistently high photosphere temperatures (e.g., \citealt{Perley+19}).  The central X-ray source responsible for the ejecta heating could be a magnetized nebula powered by the compact object (e.g., \citealt{Vurm&Metzger21}), or radiation from the inner regions of a super-Eddington accretion disk (e.g., \citealt{Sadowski&Narayan15}) or collimated jet interacting with the surrounding stellar material (e.g., \citealt{Gottlieb+22}).  The reprocessing becomes less efficient with time as the polar ejecta clears out and the dense equatorial shell takes over as the dominant reprocessing region, itself carrying the photosphere after the observed transition to lower spectral velocities.  Emission from a torus-like geometry viewed out of the equatorial plane can also account for the asymmetric shape of the spectral lines \citep{Margutti+19}.  

The broadened Fe-line fluorescence and Compton hump features can be imprinted by the propagation of a comparatively featureless (e.g., power-law) intrinsic spectrum of the central X-ray source through the fast polar shell.  The transient nature and timescale of the Compton hump feature ($t_{\rm h} \sim 10$ days) is dictated by the brief epoch over which the gas column through the polar ejecta is sufficient to generate this feature (requiring a Thomson optical depth $\tau_{\rm T} \equiv M_{\rm fast}\kappa_{\rm es}/4\pi(v_{\rm fast}t_{\rm h})^{2} \sim few$, where $\kappa_{\rm es}$ is the electron scattering opacity), supporting a similar value of $M_{\rm fast} \sim 0.1 M_{\odot}$ indicated from the optical rise time.  The total kinetic energy contained within the fast component of the AT2018cow outflow $\gtrsim M_{\rm fast}v_{\rm fast}^{2}/2 \sim 10^{51}$ erg is similar to or exceeds that of ordinary SNe.  A large quantity of radioactive $^{56}$Ni in the ejecta is disfavored by the lack of a second peak in the optical light curve at late times (e.g., \citealt{Perley+19}), likely requiring $M_{\rm Ni} \lesssim 0.1M_{\odot}$ (e.g., \citealt{Margutti+19}; however, see \citealt{Leung+20}).

Another feature of AT2018cow is its bright radio and millimeter synchrotron emission \citep{Ho+19,Margutti+19,Nayana&Chandra21}.  This radiation arises from the shock interaction between the fast polar ejecta (velocity $\approx v_{\rm fast} \sim 0.1-0.2$ c) and CSM of density $n \gtrsim 10^{5}$ cm$^{-3}$ on radial scales $\sim 10^{16}$ cm.  The CSM mass on this scale is orders of magnitude higher than predicted for a massive star wind, indicating either an outer extension of the dense CSM implicated on smaller radial scales by the narrow He emission line features \citep{Fox&Smith19} or a distinct physical origin.  After about a month of evolution, the radio emission exhibits a steep temporal drop (e.g., \citealt{Ho+19}).  Some of the atypical properties of the radio emission compared to radio SNe (e.g., the steep spectral shape, rapid post-peak decay) can be understood if the radiating electrons possess a relativistic Maxwellian energy distribution instead of the usually assumed non-thermal power-law spectrum \citep{Margalit&Quataert21,Ho+21}.   

Over the past few years, additional LFBOTs have been discovered which show qualitatively similar multi-wavelength properties to AT2018cow.  CSS161010 \citep{Coppejans+20} and AT2018lug (``Koala"; \citealt{Ho+20}) exhibited broadly similar optical and radio emission to AT2018cow, but with even higher peak ejecta speeds $v \gtrsim 0.55c$ and $v \gtrsim 0.38 c$, respectively.  AT2020xnd showed broadly similar optical, X-ray and mm radio emission to AT2018cow, including the abrupt decrease in the X-ray luminosity around 1 month after the start of the explosion \citep{Bright+21,Perley+21,Ho+21}.  The radio emission from AT2020xnd also showed evidence for an steep cut-off $n \propto r^{-3}$  in the CSM density at $r \sim 3\times 10^{16}$ cm \citep{Bright+21,Ho+21}, incompatible with that of a steady pre-explosion wind ($n \propto r^{-2}$).  AT2020mrf exhibited optical and radio emission similar to AT2018cow, but showed variable X-ray emission a factor $\gtrsim 200$ times more luminous than AT2018cow at a comparable epoch \citep{Yao+21}.  In the X-ray reprocessing picture, a higher X-ray luminosity could result from an intrinsically more powerful central engine and/or a more pole-on viewing angle in the case of geometric beaming.

Another important clue to the nature of LFBOTs is their preference to occur in low-mass starburst galaxies with moderate metal enrichment (e.g., \citealt{Coppejans+20,Ho+19,Lyman+20,Yao+21}), environments similar to those which host other engine-powered transients: long-duration gamma-ray bursts (LGRB) and Type I SLSNe (SLSN-I; e.g., \citealt{Inserra19}), the progenitors of which include the most massive stars that may require low metallicities for their formation (however, see \citealt{Michalowski+19}).  \citet{Sun+22} recently reported the detection of a temporally-stable luminous UV source ($L > 10^{40}$ erg s$^{-1}$; $T_{\rm eff} > 4\times 10^{4}$ K) with $H\alpha$ emission features spatially coincident with AT2018cow, at two epochs taken around 2-3 years after the explosion.  If interpreted as either a single-star companion or star cluster containing the progenitor system, this also favors AT2018cow being associated with very massive stars (\citealt{Sun+22}; although we explore an alternative source of transient late-time UV emission in Sec.~\ref{sec:engine}).

In summary, LFBOTs arise from a highly luminous non-thermal X-ray/$\gamma$-ray source embedded inside an extremely aspherical, potentially $^{56}$Ni-deficient ejecta shell with a wide range of velocities: a fast polar component of velocity $v_{\rm fast} \sim 0.1-0.5 c$, mass $M_{\rm fast} \sim 0.1M_{\odot}$ and kinetic energy $\gtrsim 10^{51}$ erg, as well a slower equatorial component of velocity $v_{\rm slow} \sim 3000$ km s$^{-1}$ of less well-constrained mass and energy.  The explosion environment contains H-depleted but not H-free, slower expanding $\ll v_{\rm slow}$ material extending from radii $\sim 10^{14}$ cm to $\sim 10^{16}$ cm, with a sharp outer cut-off.  The host galaxy similarities with SLSNe/LGRBs and persistent optical source imply a likely association with very massive stars. 

\subsection{Progenitor Models}

A wide range of LFBOT progenitor models of varying detail exist in the literature, including: the successful core-collapse SN explosion of a rapidly-rotating massive star with a low total ejecta mass giving birth to a central engine, such as a millisecond period neutron star (NS) or black hole (BH; \citealt{Prentice+18,Perley+19,Margutti+19,Gottlieb+22}); an initially ``failed'' SN which nevertheless produces an accreting BH and mass ejection via accretion disk-winds \citep{Quataert+19,Perley+19,Margutti+19,Antoni&Quataert22}; the tidal disruption of a star by an intermediate-mass BH \citep{Kuin+19,Perley+19} or a stellar-mass BH in a dense stellar environment \citep{Kremer+21}; and shock interaction with a dense circumstellar outflow from the progenitor following the explosion of a H-poor star (e.g., \citealt{Fox&Smith19}) or pulsational pair instability SNe (e.g., \citealt{Leung+20}), among other possibilities to be discussed below.

\begin{table*}
\centering
\small
\caption{Progenitor Models for LFBOTs Must Confront All Observations \label{tab:models}}
\begin{tabular}{ccccccccc}
\hline
Progenitor & Extreme Ejecta  & Compact & Extended CSM  & H-Depleted & Very Massive & Low & Example \\
Model & Asphericity$^{(a)}$ & Object & ($\gtrsim 10^{16}$ cm) & CSM & Stars/Low $Z$ & $^{56}$Ni & References \\
\hline
Engine-Powered SN & $\text{\sffamily X}$ & \checkmark & ? & \checkmark & \checkmark & ? & 1 \\
Failed SN+BH disk$^{\dagger}$ & \checkmark & \checkmark  & ? & ? & ? & \checkmark & 2\\
IMBH TDE & \checkmark & \checkmark & $\text{\sffamily X}$ & ? & $\text{\sffamily X}$ & \checkmark & 3 \\
BH+Star Merger & \checkmark & \checkmark  & $\text{\sffamily X}$ & ? & $\text{\sffamily X}$ & \checkmark & 4 \\
PPISN  &  $\text{\sffamily X}$ &  ? & ? & \checkmark &  \checkmark &  $\text{\sffamily X}$ & 5 \\
Failed PPISN+BH disk$^{\dagger}$  &  \checkmark &  \checkmark & ? & \checkmark &  \checkmark &  \checkmark & Future Work \\
Failed CE + Prompt Merger  & \checkmark & \checkmark & \checkmark & $\text{\sffamily X}$ & 6 & ? & \checkmark \\
Failed CE + Delayed Merger$^{\dagger}$ & \checkmark & \checkmark & \checkmark & \checkmark & \checkmark & \checkmark & This Paper\\

\hline \\
\end{tabular}
\\
$^{\dagger}$Allowed LFBOT contenders, subjectively defined as models without a $\text{\sffamily X}$ mark in any column.\\
References: (1) \citealt{Prentice+18,Margutti+19}; (2) \citealt{Quataert+19,Perley+19,Margutti+19}; (3) \citealt{Perley+19}; (4) \citealt{Kremer+19}; (5) \citealt{Leung+21}; (6) \citealt{Soker+19,Soker19,Schroder+20}\\
$^{(a)}$As implied by the large range of ejecta velocities $\sim 0.01-0.1$ c, required across different ejecta latitudes, to simultaneously explain the fast optical rise, early escape of X-rays, and bright synchrotron radio emission (polar regions), as well as the much lower late-time spectral line widths and persistent photosphere emission requiring denser slower material (equatorial regions).
\end{table*}

As we shall argue, few if any of the these scenarios provide a satisfactory explanation for all LFBOT observations and basic inferences drawn from them (Table \ref{tab:models}).  Models that rely exclusively on CSM interaction (e.g., \citealt{Fox&Smith19,Schroder+20,Leung+21,Margalit21,Dessart+21,Pellegrino+22}) provide no explanation for the presence of a highly time-variable central non-thermal radiation source \citep{Margutti+19,Yao+21}, which is clearly manifest via the X-ray/gamma-ray emission in AT2018cow/AT2020mrf and likely requires an energetic compact object.  Such a large energy in ejecta expanding at trans-relativistic speeds (up to $\sim 0.5 c$; \citealt{Coppejans+20}) also points to a BH or NS central engine.  As does the quasi-periodic X-ray feature, which if interpreted as an orbital frequency in an accretion disk, constrains the central object mass $\lesssim 850M_{\odot}$ \citep{Pasham+21}. 

Models which invoke successful core-collapse SNe are challenged to explain the extremely asymmetric stellar ejecta (implied by the wide range of outflow velocities) compared to other engine-powered stellar explosions such as SLSN-I (e.g., \citealt{Inserra+16}) or the broad-lined Type Ic SNe that typically accompany long gamma-ray bursts (e.g., \citealt{Stevance+17}).  An intrinsically aspherical core-collapse explosion is also disfavored theoretically by the fact that stars with low envelope masses (and hence low ejecta masses) are found by modern SN simulations to explode promptly via the neutrino-driven mechanism (e.g., \citealt{Melson+15,Lentz+15}).  

The initially failed explosion of a single massive star which forms a BH but nevertheless produces a small quantity of ejecta (e.g., \citealt{Quataert+19,Antoni&Quataert22}) can more naturally explain the aspherical ejecta geometry because the ``explosion'' is driven by disk winds instead of a quasi-spherical SN shock.  However, this scenario by itself does not explain the ubiquitous presence of massive CSM extending out to $\sim 3\times 10^{16}$ cm (tidal disruption events and BH-star collisions are disfavored for the same reason).  A small fraction of massive stars do exhibit strongly enhanced mass-loss rates just prior to exploding (e.g., \citealt{Kiewe+12,Taddia+13}), but none of the existing explanations for this behavior in stripped envelope stars (e.g., \citealt{Fuller&Ro18}) predict why it should occur preferentially from the same progenitors which upon collapse create the most energetic central compact objects.  

Stated another way, the biggest challenge to any model for LFBOTs is to {\it explain the simultaneous presence of an energetic compact object and a massive and radially-extended $\sim 10^{14}-10^{16}$ cm dense medium surrounding the explosion.}  Occam's Razor dictates that two atypical properties$-$in this case, a central engine and dense CSM of a consistent radial distribution$-$are probably related.  

Pulsational pair instability supernovae (PPISN), for particular choices of the SN explosion energy and pre-explosion CSM arising from previous PPI mass-loss events, can generate fast-evolving optical light curves via shock interaction of the supernova ejecta with extended CSM on radial scales $\sim 10^{14}$ cm \citep{Woosley17,Renzo+20,Leung+21}.  However, as already noted, the coupled behavior and similar late-time luminosities of the optical and X-ray light curves in AT2018cow favors reprocessed engine-power instead of shock interaction as the dominant source of optical emission.  Furthermore, multiple PPI eruptions may be required to generate the large CSM mass covering radii $\sim 10^{14}-10^{16}$ cm (potentially requiring a fine-tuned progenitor star mass), despite a broadly similar CSM density field now inferred to characterize multiple members of the LFBOT class \citep{Ho+19,Bright+21,Ho+21}.  The low polar ejecta mass $M_{\rm fast} \lesssim 0.1M_{\odot}$ which enables soft X-rays to escape from the central engine \citep{Margutti+19}, is also in tension with the large ejecta masses predicted from the successful explosion of a $\sim 40M_{\odot}$ PPISN progenitor.  

The core-collapse of stars which undergo PPI may not give rise to successful energetic SN explosions due to their massive Fe cores and the large gravitational binding energy of their envelopes (e.g., \citealt{Powell+21,Rahman+21}).  An alternative, potentially more promising PPISN scenario for LFBOTs would therefore invoke an initially failed neutrino-driven explosion giving rise to prompt BH formation, followed by a delayed wind-driven explosion once the outer stellar layers form an accretion disk around the BH (similar to the scenario explored in \citealt{Siegel+21} in the context of even more massive stars above the pair-instability mass gap).  This could better explain the low ejecta mass and extreme asymmetry of the ejecta (similar to the failed SNe models described above) as well as the presence of a central engine.  PPISN progenitors may be challenged to retain enough angular momentum to generate an accretion disk upon collapse, particularly in face of stellar winds during previous stages of evolution (as necessary to remove most of the H envelope).  Nevertheless, failed PPISNe giving rise to hyper-accreting BH engines, should be considered as LFBOT progenitors in future work.

\subsection{This Paper: The Delayed Merger Scenario}

The present paper focuses on a second promising class of models, which generally begins with common envelope (CE) interaction following unstable mass-transfer from a giant star onto a BH or NS binary companion (\citealt{Soker+19,Uno&Maeda20,Schroder+20}; see also \citealt{Chevalier12}).  After the BH/NS plunges into the envelope of the giant and spirals towards its center, the gravitational energy released is generally expected to unbind the stellar envelope (e.g., \citealt{MacLeod+18,Law-Smith+20}; however note that there exist ways to tighten the binary through stable mass-transfer without a CE event; e.g., \citealt{vandenHeuvel17,Pavlovskii+17,Neijssel+19,Klencki+21,Marchant+21,vanSon+21}).  If the envelope cannot be removed (i.e., the CE is a ``failure''), past works envision that the BH/NS will enter the He-rich core of the giant, triggering a high accretion rate onto the BH/NS and powering a ``merger-driven'' explosion (e.g., \citealt{Fryer&Woosley98,Chevalier12,Soker+19,Schroder+20}).  A failed CE and prompt explosion model was recently invoked by \citet{Dong+21} to explain the radio transient VT J121001+495647 and its high-energy precursor.  \citet{Soker22} propose an alternative version of this scenario (``Common Envelope Jets Supernova'' model), in which an extended phase of mass-loss from the giant preceding the first common envelope event creates a radially extended CSM into which the subsequent merger-driven jetted explosion interacts.

Though avoiding many of the pitfalls of other LFBOT models, the failed CE scenario is subject to its own challenges, at least as it has been presented in the literature thus far.  Firstly, there is the theoretical issue of how the central ``explosion'' is typically modeled.  The physical picture of the BH/NS ``entering'' the He core and accreting at some prescribed rate based on its density (e.g., Bondi-Hoyle), is not realistic in the presence of substantial angular momentum.  The process is probably better described as the He core being tidally disrupted and then accreted by the BH/NS through a rotationally-supported disk, powering the explosion via disk winds (e.g., \citealt{MacFadyen&Woosley99}).\footnote{For similar reasons, \citet{ThorneZytkow77} objects appear challenging to form through such a scenario.}  The relevant duration of the central engine's activity is then set by the viscous accretion time of the disk, rather than the density of the undisrupted core (though the two are intimately related because the core's tidal radius determines the characteristic accretion disk size).  Similar accretion-powered transients have been studied in the context of the merger of a white dwarf with a BH/NS (e.g., \citealt{Metzger12,Fernandez&Metzger13,Margalit&Metzger16,Zenati19b,Fernandez+19,Zenati+20,Bobrick+21}), the knowledge from which can be applied to quantify the kinetic and nucleosynthetic output of the heretofore invoked merger-driven ``explosions''. 

Another challenge to CE models is the mass and composition of the CSM predicted from a CE phase.   Very massive progenitor stars are suggested by the host galaxy properties of LFBOTs and, at least in the case of BH accretors, failed CE events should preferentially occur for massive donor stars $\gtrsim 20-40 M_{\odot}$ (e.g., \citealt{Kruckow+16,Schroder+20,Lau+22}) with similarly massive envelopes.  However, the fast optical rise times and the prompt escape of soft X-rays from LFBOTs require the polar ejecta at radii $r \lesssim v_{\rm fast}t_{\rm pk} \lesssim 10^{15}$ cm to possess a low mass $\lesssim 0.1-1 M_{\odot}$.  More CSM mass would be allowed in the equatorial plane (and CE ejecta is indeed focused in the binary plane; \citealt{Soker+19,Schroder+20}) but it is still difficult to reconcile the presence of up to tens of solar masses of hydrogen with the observations.  In particular, the H-rich CE debris present on radial scales $\sim 10^{13}-10^{15}$ cm \citep{Soker+19,Schroder+20} would likely give rise to prominent hydrogen features in the transient spectra (e.g., analogous to Type II SLSNe or Type IIn SNe, as predicted in precisely this scenario by \citealt{Chevalier12}), inconsistent with the H-depleted Type Ibn or transitional Ibn/IIn-like spectra of AT2018cow (e.g., \citealt{Fox&Smith19}), for which at most a few tenths of a solar mass of hydrogen is likely to be present (e.g., \citealt{Dessart+21}).

As we shall argue, the lack of massive H-rich CSM can be understood if the WR-BH/NS merger is substantially delayed (e.g. by centuries or longer) following the initial CE ejection or stable Roche Lobe Overflow (RLOF) mass-transfer phase responsible for tightening the binary.  Such a delay could occur, for example, due to gradual angular momentum extraction by a relic rotating envelope or circumbinary disk left over from the original mass-transfer/CE phase (as proposed by \citealt{Kashi&Soker11} in a different context; see also \citealt{Ivanova11,Fragos+19,Lau+22} in terms of CE work suggesting a significant fraction of the envelope remains bound). Depending on system-to-system variations in the properties of immediate post-CE binary, its relic disk, and the evolutionary state of the progenitor star prior to the original mass-transfer or CE event, such delayed merger events could manifest as a diverse range of transient properties.  Of particular relevance in this context is the possible connection between LFBOTs and Type Ibn/Icn SNe \citep{Dessart+21,Perley+22,Gal-Yam+22}.  The latter class of rare fast-evolving transients share many properties with LFBOTs, including similar optical light curves and spectroscopic evidence for shock interaction between the ejecta and slow aspherical H-poor material (e.g., \citep{Foley+07,Pastorello+08,Dessart+21,Gal-Yam+22,Perley+22}; see \citet{Smith17} for a review).

\begin{figure*}
    \centering
    \includegraphics[width=0.34\textwidth]{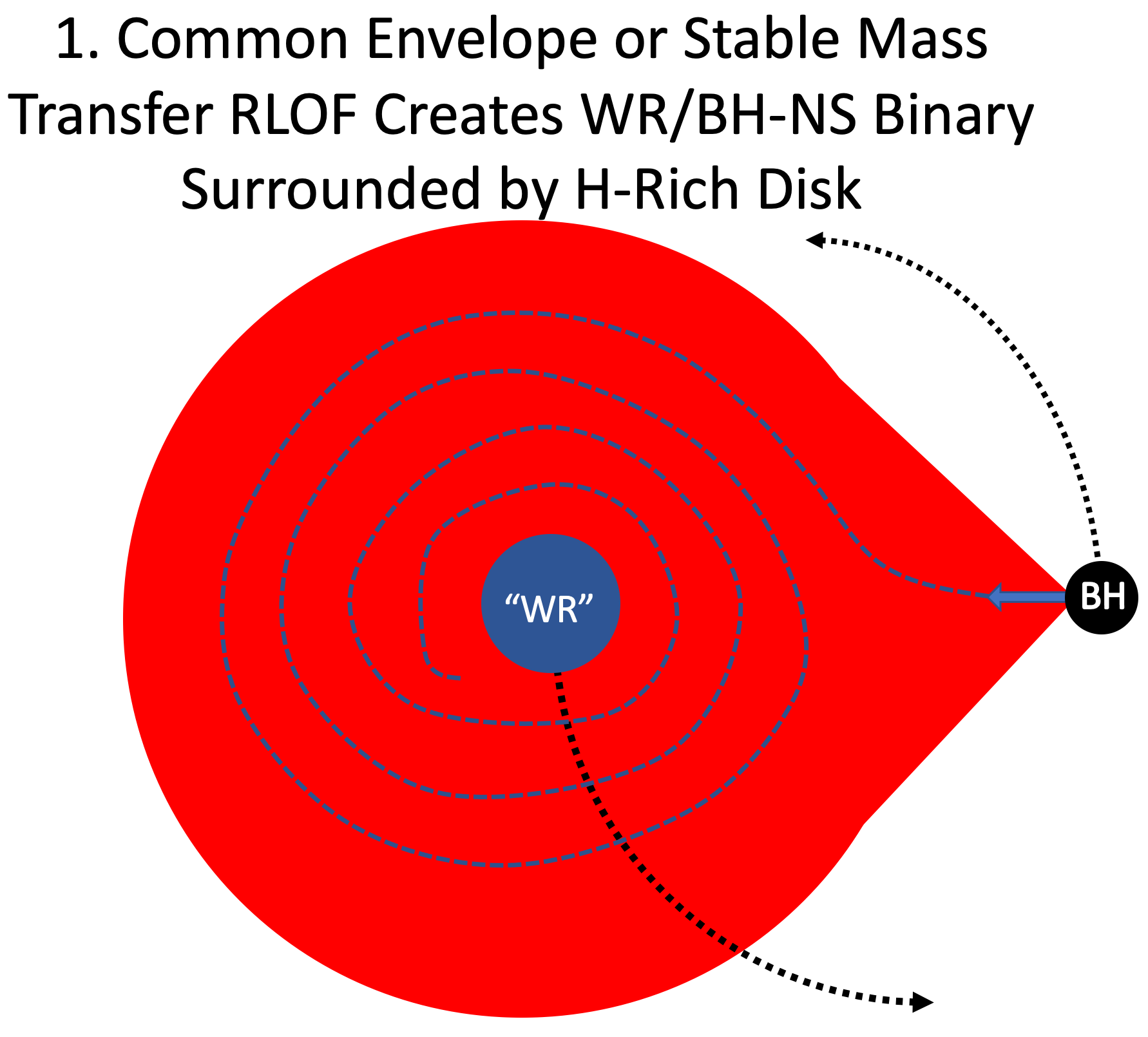}
    \includegraphics[width=0.55\textwidth]{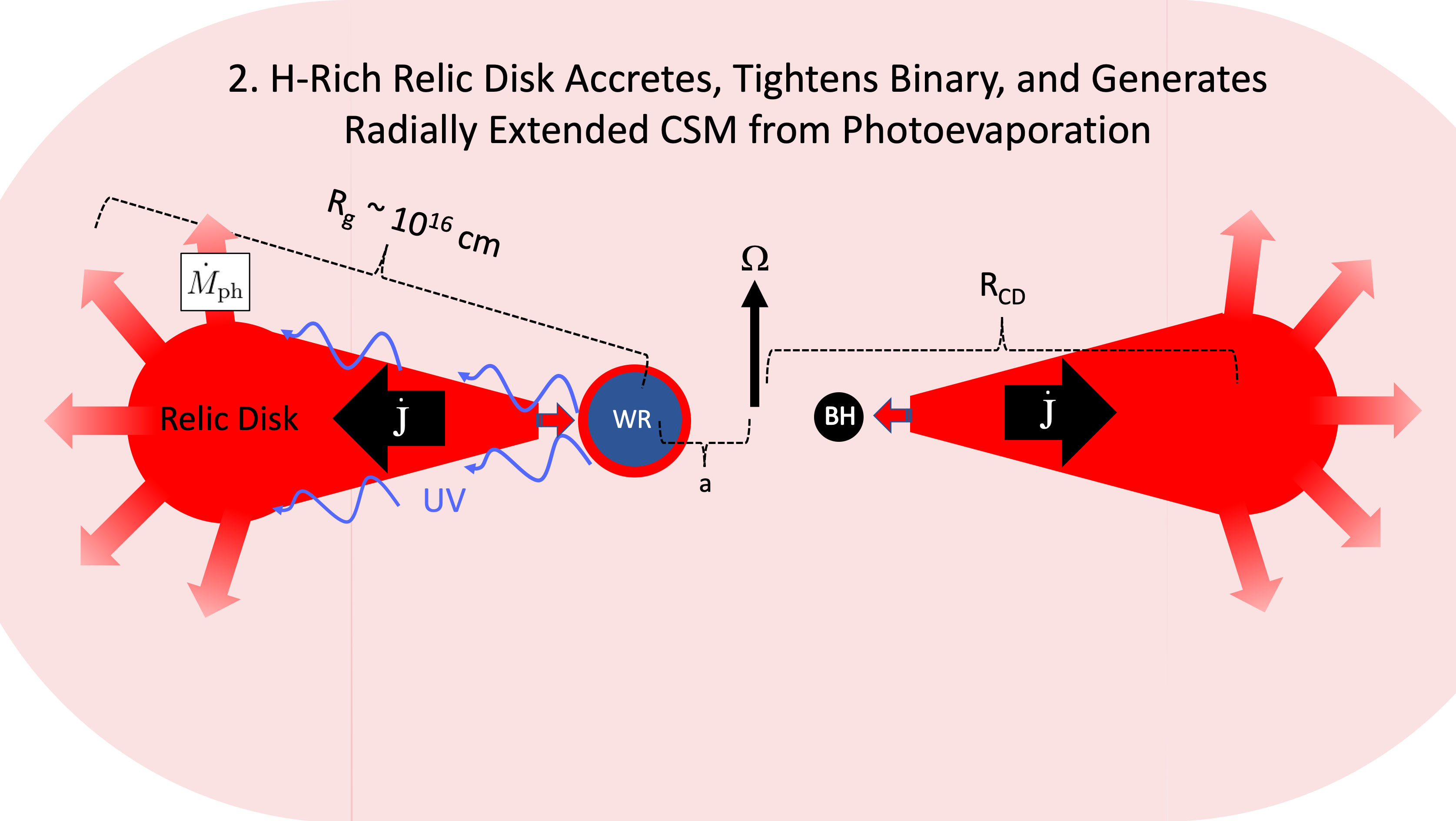}
    \includegraphics[width=0.6\textwidth]{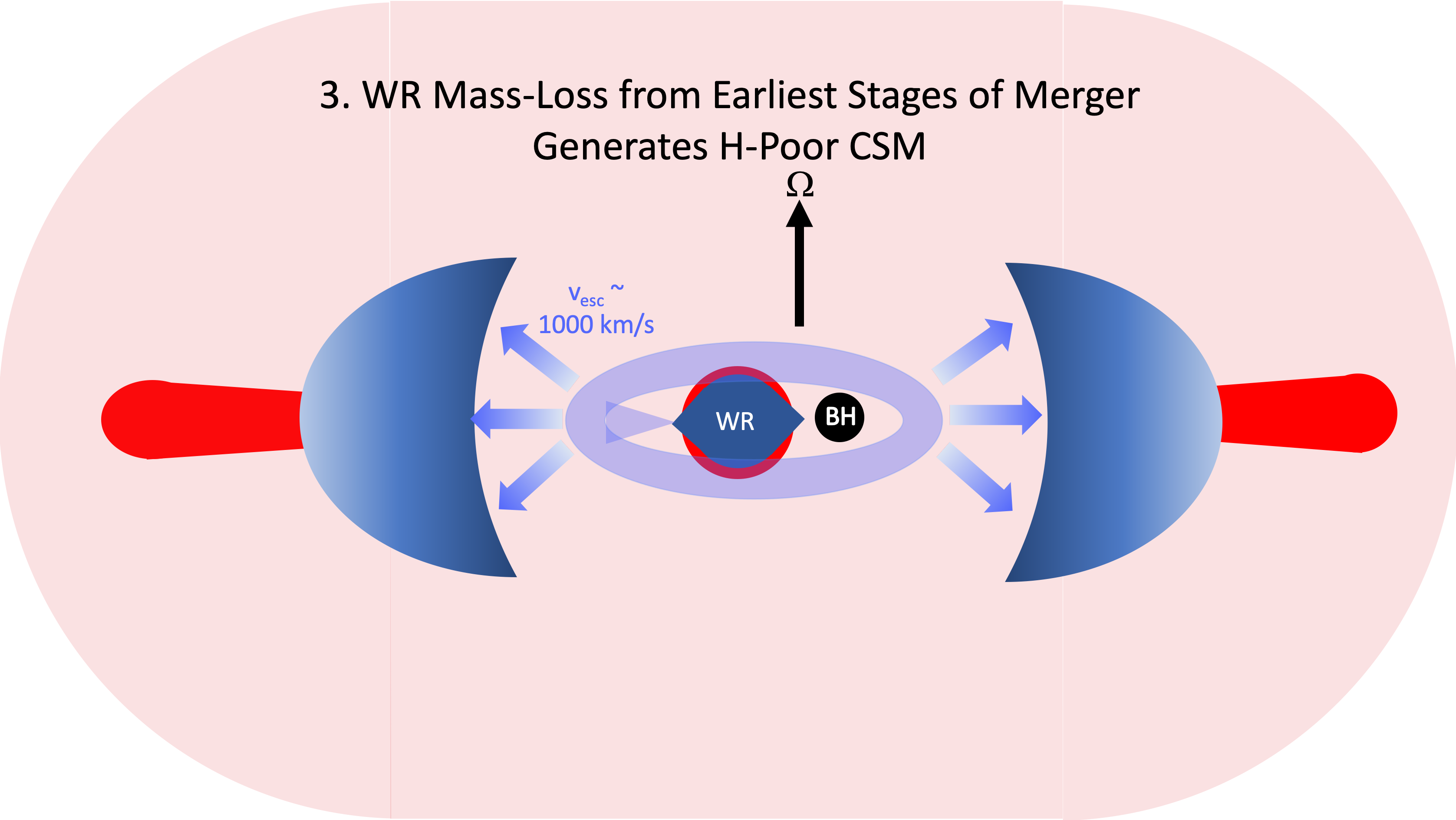}
    \includegraphics[width=0.8\textwidth]{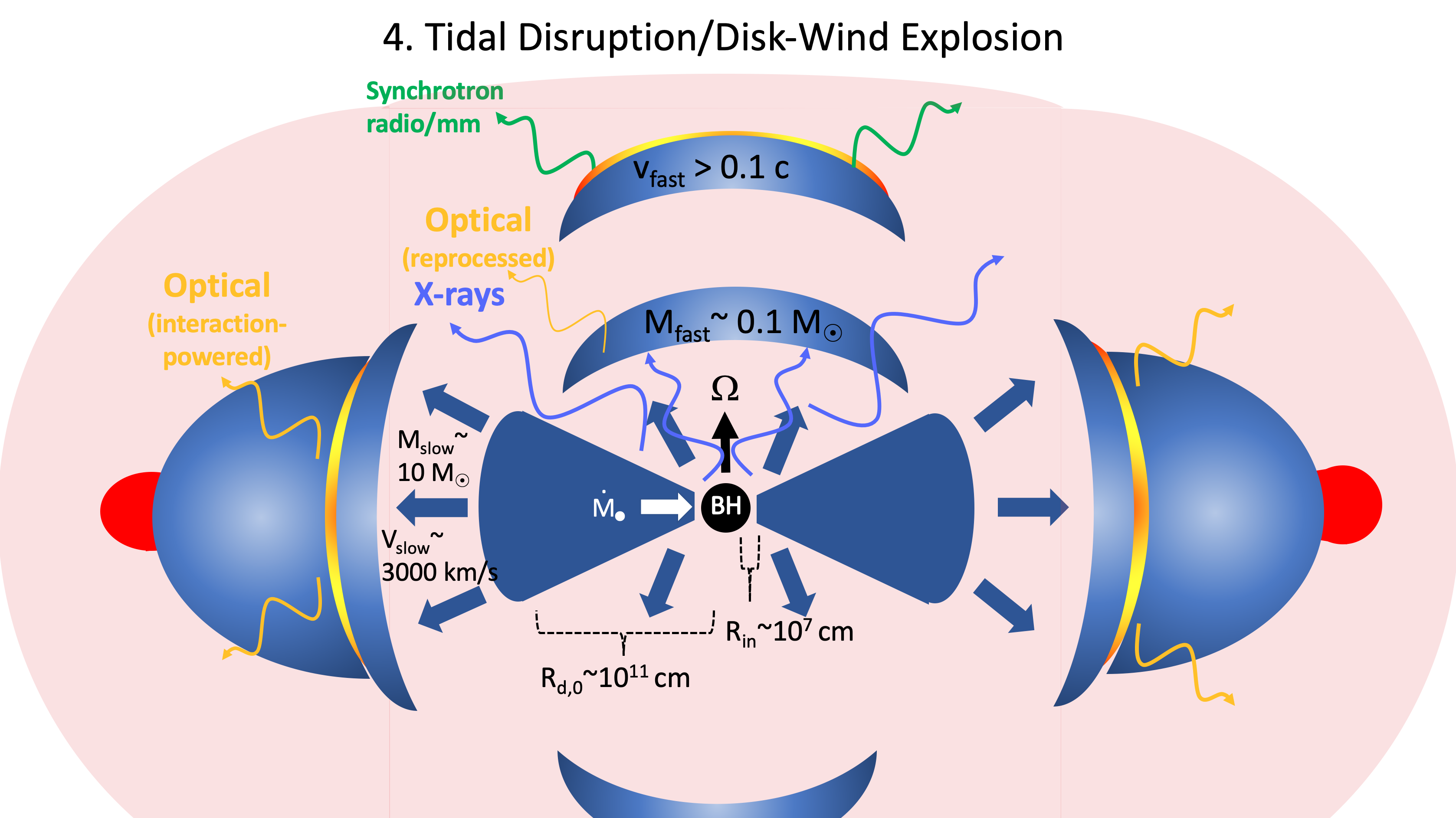}
    \caption{Timeline of LFBOT model, where red and blue colors refer to hydrogen-rich and hydrogen-poor material, respectively.  (1) CE event and/or stable mass-transfer removes H-envelope, creating tight WR-BH/NS binary; (2) Leftover H-rich disk spreads outwards, extracting angular momentum from WR-BH/NS binary and driving it to merge; Photo-ionization heating from the WR star drives an outflow from the disk radii $R_{\rm g} \sim 10^{16}$ cm (Eq.~\ref{eq:Rg}); (3) WR star undergoes RLOF onto the BH/NS, leading to unstable mass transfer.  Mass loss from the WR star from the $L_2$ point during early stages of the runaway mass-transfer process generates H-depleted CSM expanding at $\sim 10^{3}$ km s$^{-1}$ (Eq.~\ref{eq:vL2}), which collides with the pre-existing H-rich disk; (4) WR star is tidally disrupted, creating a massive disk surrounding the BH/NS.  Hyper-accretion generates fast and slow outflows from the inner and outer radii in the disk, respectively.  The optical luminosity is powered by (a) reprocessing of X-rays from the inner accretion flow or jet by the fast disk-wind ejecta; and (b) shock interaction between the equatorial disk-wind ejecta and the pre-merger CSM.  X-rays which escape the polar funnel and avoid reprocessing can be directly observed but may be geometrically beamed along the rotational axis. Radio/mm emission is powered by the interaction of the fast ejecta with the evaporated disk-wind outflows on larger scales.}
    \label{fig:cartoon}
\end{figure*}

This paper is organized as follows.  Section \ref{sec:model} details a physical model leading to the merger of a WR-BH/NS binary.  In Section \ref{sec:transient} we outline the transient electromagnetic emission from the merger and present a toy model for the optical/X-ray light curves of LFBOTs.  Section \ref{sec:unification} describes the implications of our findings for a possible unification of LFBOTs and Type Ibn/Icn into a common framework with a continuum of properties controlled primarily by a single variable: the post-CE merger delay time.  In Section \ref{sec:conclusions} we summarize our conclusions.  Figure \ref{fig:cartoon} schematically illustrates the stages of the envisioned model.

\section{Mergers of WR-BH/NS Binaries}
\label{sec:model}

This section is organized as follows.  We begin in Section~\ref{sec:together} with a brief overview of the CE phase required to create a WR-BH/NS binary and identify the most promising physical mechanism for bringing the binary into Roche Lobe contact to initiate the merger (Sec.~\ref{sec:delayed}) and prefacing the implications for a dense gaseous medium surrounding the binary at the time of the merger.  Section \ref{sec:merger} describes the merger process, including the tidal disruption of the WR star and the accretion of its debris onto the central BH/NS (Sec.~\ref{sec:accretion}) and the properties of the resulting disk-wind outflows (Sec.~\ref{sec:outflows}, \ref{sec:burning}) and innermost accretion flow luminosity (Sec.~\ref{sec:engine}).

\subsection{Path to the Merger}
\label{sec:together}

\subsubsection{Dynamical Common Envelope Phase}

\label{sec:CE}

Similar to earlier works (e.g., \citealt{Soker+19,Soker19,Schroder+20}), our envisioned scenario begins with a binary composed of a massive star $\gtrsim 20-40M_{\odot}$ orbiting a BH or NS remnant of mass $M_{\bullet} \sim 1-20 M_{\odot}$, the latter the product of an earlier core-collapse event.  For concreteness we focus on a BH companion; however, some of the considerations to follow would be qualitatively similar for a NS companion.  Massive star binaries are most consistent with the host galaxy demographics of LFBOTs and Type Icn SNe (e.g., \citealt{Coppejans+20,Yao+21,Perley+22}) and its stellar environment \citep{Sun+22}.  As the star evolves off the main sequence, its radius expands and eventually undergoes RLOF onto the BH/NS companion.  

As a fiducial case, we consider that by this point the donor star has begun He core burning and is similar to a Wolf-Rayet\footnote{Throughout this paper, we use the term ``Wolf-Rayet star'' instead of ``massive He core'' to connect more closely to the literature proposing WR stars as the progenitors of Type Ibn/Icn (e.g., \citealt{Gal-Yam+22,Perley+22}).  We note that not all WR stars possess large He cores (e.g., WO stars), nor would all He-burning stars be classified as WR stars (e.g., transparent wind stripped-envelope stars).} star (WR) in terms of its structure; however, a qualitatively similar outcome to what follows may result if the star is lower mass or if its core has not yet ignited He.  Depending on binary mass ratio and the depth of the surface convection zone, the mass-transfer process onto the BH/NS can evolve to dynamical instability (e.g., \citealt{Pavlovskii+17}).  The eventual outcome is typically for the BH/NS to lose co-rotation with and plunge into the envelope of the donor star, instigating a common envelope (CE) event (e.g., \citealt{Taam+78,Iben&Livio93,Armitage&Livio00,Ivanova+13,DeMarco&Izzard17,Lau+22}).

Hydrodynamical simulations of the CE phase reveal that the accretor plunges within the donor's envelope, spiraling into its center on a few dynamical timescales, approximately equal to the orbital period at the donor's surface (e.g., \citealt{MacLeod+18}).
The outcome of a given CE event, particularly the fraction of the primary's envelope which is unbound, and the resulting final separation between the BH/NS remnant and the evolved WR stellar core, are uncertain theoretically and remain areas of active research (e.g., \citealt{Ivanova+13,Wilson&Nordhaus19,Moreno+21,Lau+22}), with implications for e.g.~the rates of binary compact object mergers  (e.g.~\citealt{Belczynski+16,Tauris+17,Vigna-Gomez+18,Broekgaarden+21}).

In some systems the CE may be replaced or preceded by a phase of thermal-timescale stable mass-transfer (e.g., \citealt{Pavlovskii+17,vandenHeuvel17,Klencki+21,Marchant+21,vanSon+21,Bavera+21,Gallegos-Garcia+21}), which also has the effect of at least partially tightening the binary on a relatively short timescale (and may leave a relic gaseous disk due to RLOF spillover from the binary during the mass-transfer process; see below).

Here, we gloss over these details and assume that the CE or mass-transfer phase promptly removes most of the hydrogen envelope from the binary, which expands away from the system at a velocity of typically $v_{\rm CE}\sim $ tens to hundreds of km s$^{-1}$ (depending on the evolutionary state of the donor and the mechanism of mass-loss), leaving the WR star core of the evolved star orbiting the BH/NS remnant at a binary separation $a_0$ much smaller than the original donor star radius.  However, contrary to previous works on explosive CE-transients (e.g., \citealt{Soker+19,Schroder+20}), we do not assume the BH/NS immediately merges with the stellar core.  Events of the latter type, as a result of the lingering presence of the slowly evacuating massive giant envelope, may give rise to a luminous Type II/IIn SN (\citealt{Chevalier12,Dong+21}) or a Type Ib or Ic SN that transforms into a Type IIn (e.g., \citealt{Tinyanont+16,Chen+18,Thomas+22}).  Such prompt core merger explosions are not however likely to account for the strongly H-depleted CSM which characterizes LFBOTs (e.g., \citealt{Fox&Smith19}) much less Type Ibn/Icn SNe (e.g., \citealt{Dessart+21}).

\subsection{Delayed Binary Merger}
\label{sec:delayed}

Rather than a dynamical timescale merger driven by the mass-transfer/CE event itself, we consider processes that may lead to a WR-BH/NS merger over a much longer period $\gtrsim$ hundreds or thousands of years.

To start, consider processes that act on the binary absent any surrounding gas.  Orbital decay due to gravitational waves occurs on a timescale,
\be t_{\rm GW} = \frac{a}{\dot{a}} \approx \frac{5c^{5}a^{4}}{128G^{3}M_{\star}^{3}} \approx 5\times 10^{6}{\rm yr}\,M_{\star,10}^{-0.7}\left(\frac{a_0}{2.6R_{\star}}\right)^{4},
\ee
where we have assumed an equal mass binary ($M_{\bullet} \approx M_{\star}$) and have normalized the post-CE binary semi-major axis $a = a_0$ to the value at which the WR star undergoes RLOF onto the BH/NS (see Eq.~\ref{eq:aRLOF} below).  Here, $M_{\star} = 10M_{\star,10}M_{\odot}$ is the mass of the WR star, which is related to its radius approximately as (e.g., \citealt{Schaerer&Maeder92})
\be
R_{\star} \approx 0.8R_{\odot}M_{\star,10}^{0.58}
\label{eq:Rstar}.
\ee
Since $t_{\rm GW}$ is much longer than the nuclear evolution time $\lesssim 10^{5}$ yr of the WR star, the latter will evolve as being effectively single, eventually undergoing its own core-collapse and (potentially) exploding as a stripped-envelope Type Ib/Ic SN, creating a double compact object binary system (e.g., \citealt{VignaGomez+21}).
Accretion of the SN ejecta by the BH/NS during this process could potentially power a high energy transient similar to a GRB (e.g., \citealt{Rueda&Ruffini12,Fryer+14}).  However, the long delay $\gtrsim 10^{5}$ yr until the explosion ensures that any debris from the earlier CE event will have completely cleared out of the system, contrary to the CSM interaction in LFBOTs and Ibn/Icn SNe.  The conclusion that gravitational waves alone are unlikely to lead to a merger is further strengthened by the competing effect of binary {\it widening} due to wind mass-loss from the WR star (however, see \citealt{Detmers+08}).

Another process may act to drive together the WR-BH/NS binary on a timescale intermediate between that of the dynamical-timescale of the CE and the core's nuclear evolution.  If a small fraction of the CE envelope (e.g., \citealt{Reichardt+19}) remains bound to the surviving binary in a rotationally supported disk, this disk can extract angular momentum through viscous torques (e.g., \citealt{Pringle91,Taam&Spruit01}) in a process akin to Type II planetary migration, driving the binary to merge with some delay after the CE (e.g., \citealt{Kashi&Soker11,Soker13}).  

A Keplerian gaseous disk of mass $M_{\rm CD} \ll M_{\rm bin}$ can extract sufficient angular momentum to drive a binary of mass $M_{\rm bin} = M_{\star} + M_{\bullet}$ and semi-major axis $a_0$ to merge by spreading viscously outwards to a radius $R_{\rm CD} \sim a_0(M_{\rm bin}/M_{\rm CD})^{2}$ as dictated by conservation of angular momentum $J_{\rm CD} \simeq M_{\rm CD}(GM_{\rm bin}R_{\rm CD})^{1/2} = J_{\rm bin} \sim M_{\rm bin}(GM_{\rm bin}a_0)^{1/2}$.  The spreading process and hence the merger occurs roughly on the viscous timescale of the disk near its outer edge $r \sim R_{\rm CD}$ (e.g., \citealt{Frank+02}),
\begin{eqnarray}
&& t_{\rm visc}^{\rm CD} \sim \left.\frac{r^{2}}{\nu}\right|_{R_{\rm CD}} \sim \frac{1}{\alpha}\frac{1}{\theta^{2}}\left(\frac{R_{\rm CD}^{3}}{GM_{\rm bin}}\right)^{1/2} \nonumber \\
 &\approx& 140\,{\rm yr}\, \alpha_{0.1}^{-1}\theta_{0.33}^{-2}\left(\frac{a_0}{30R_{\odot}}\right)^{3/2}\left(\frac{M_{\rm bin}}{30M_{\odot}}\right)^{-1/2}\left(\frac{M_{\rm CD}}{0.1M_{\rm bin}}\right)^{-3},  \nonumber \\ \label{eq:tvisc}
\end{eqnarray}
where $\nu = \alpha c_{\rm s}H = \alpha r^{2}\Omega_{\rm K}\theta^{2}$ is the effective viscosity and $\theta = H/r$ the aspect ratio of the disk of vertical thickness $H$; $\Omega_{\rm K} \equiv (GM_{\rm bin}/r^{3})^{1/2}$ is the Keplerian orbital frequency; $c_{\rm s} \approx H\Omega_{\rm K}$ is the sound speed; and $\alpha = 0.1\alpha_{0.1}$ is the viscosity parameter \citep{Shakura&Sunyaev73} scaled to a typical value for turbulence generated by the magneto-rotational instability (MRI; \citealt{Balbus&Hawley98}).  The characteristic accretion/decretion rate $\dot{M}_{\rm CD} \sim M_{\rm CD}/t_{\rm visc}^{\rm CD}$ is then given by,
\begin{eqnarray}
&&\frac{\dot{M}_{\rm CD}}{\dot{M}_{\rm Edd}} \approx 6\times 10^{-3} \alpha_{0.1}\theta_{0.33}^{2}\times \nonumber \\
&&\left(\frac{a_0}{30R_{\odot}}\right)^{-5/2}\left(\frac{M_{\rm bin}}{30M_{\odot}}\right)^{3/2}\left(\frac{M_{\rm CD}}{0.1M_{\rm bin}}\right)^{6},
\label{eq:MdotCD}
\end{eqnarray}
where $\dot{M}_{\rm Edd} \equiv L_{\rm Edd} R_{\rm CD}/(GM_{\rm bin})$ is the local Eddington accretion rate and $L_{\rm Edd} \simeq 5\times 10^{39}(M_{\rm bin}/30M_{\odot})$ erg s$^{-1}$ is the electron-scattering Eddington luminosity.  

The above estimates can only be taken as order-of magnitude.  The strength of the torque exerted on the binary by the gaseous orbiting disk (and even its sign) remain areas of active research (e.g., \citealt{Tiede+20,Munoz+20,Dittman&Ryan22,Penzlin+22,Dempsey+22}); a more precise estimate of the binary hardening time will depend on the dependence of the disk-binary torque on the binary mass ratio, disk thickness, and eccentricity (e.g., \citealt{Duffell+20,DOrazio+21}), and how these evolve as the disk spreads. 

For massive disks $M_{\rm CD} \gtrsim 0.2M_{\rm bin}$ and/or tight post-CE binaries $a_0 \lesssim 30 R_{\odot}$, Eq.~\ref{eq:MdotCD} predicts that the circumbinary disk accretion rate is super-Eddington $\dot{M}_{\rm CB} \gtrsim \dot{M}_{\rm Edd}$, justifying our assumption of a geometrically-thick disk ($\theta \sim 1$) and implying a comparatively short merger delay of $\sim t_{\rm visc}^{\rm CD} \sim 10-1000$ yr.  This delay may nevertheless be long enough to evacuate debris from the CE or thermal mass-transfer phase (of expansion velocity $v_{\rm CE} \gtrsim 100$ km s$^{-1}$) to radii $\sim v_{\rm CE}t_{\rm visc}^{\rm CD} \gtrsim 10^{16}-10^{17}$ cm (we shall return to the implications for CSM interaction in Sec.~\ref{sec:radio}).  

For lower mass disks $M_{\rm CB} \ll 0.2 M_{\rm bin}$ and/or larger binary separations $a_0 \gtrsim 30R_{\odot}$, the circumbinary accretion rate will be sub-Eddington, the disk thinner $\theta \ll 1$, and the merger time much longer $t_{\rm visc}^{\rm CD} \gg 10^{3}$ yr.  In this case the unbound ejecta from the CE or binary mass-transfer phase will have completely cleared out, to large radii $\gg 10^{17}$ cm by the time of the WR-BH/NS merger.       

In both cases, enough mass can remain in the circumbinary disk itself $M_{\rm CD} \sim 0.1-1M_{\odot}$ on radial scales $\sim R_{\rm CD} \gtrsim (100-1000) a_0 \gtrsim 10^{14}-10^{15}$ cm to leave some hydrogen in the system and otherwise influence the merger-driven transient (Sec.~\ref{sec:transient}).  Outflows from the circumbinary disk, driven by photoevaporation by the WR star, may dominate the CSM environment surrounding the binary on the largest scales $\gtrsim 10^{16}$ cm (Sec.~\ref{sec:CSM}).  

As discussed above, the merger delay time (on which the type of supernova-like transient to follow depends; Sec.~\ref{sec:unification}) depends sensitively on the initial mass $M_{\rm CD}$ and size $a_{\rm 0}$ of the remnant CE disk, which could vary systematically with the progenitor binary properties.  For example, the CE simulations by \citet{Lau+22} predict that mergers involving BH companions leave more extended relic envelopes (larger $a_0$) than those with NS companions; if representative, this suggests WR-BH mergers may preferentially occur with longer delays than WR-NS mergers.  Likewise, insofar as mergers involving more massive donor stars (with more massive envelopes and greater gravitational binding energies) are less likely to result in successful CE events (e.g., \citealt{Kruckow+16}) and hence would be expected to generate more massive leftover disks (larger $M_{\rm CD}$), the merger delay time might also be expected to decrease with increasing donor mass.  A systematic study of long-term evolution of the relic disk, using initial conditions motivated by the outcome of CE simulations finding `failed' envelope removal, accounting also for mass-loss from the leftover disk due to radiation and wind feedback from the stripped star (Sec.~\ref{sec:binarydiskoutflows}), is needed to better quantify how the range of possible outcomes map to the properties of the original merging binary. 

\subsection{Merger and Accretion-Powered Explosion}
\label{sec:merger}

Regardless of what process ultimately brings the binary together, the WR star will undergo RLOF onto the BH/NS once the binary orbital separation shrinks to the value \citep{Eggleton83}
\be
a_{\rm RLOF} \approx R_{\star}\frac{0.6q^{2/3} +{\rm ln}(1+q^{1/3})}{0.49q^{2/3}} \underset{q \simeq 1}\approx 2.6 R_{\star},
\label{eq:aRLOF}
\ee
where $q \equiv M_{\star}/M_{\bullet}$.  The orbital period of the binary at this point,
\be
P_{\rm orb} = 2\pi \left(\frac{a_{\rm RLOF}^{3}}{GM_{\rm bin}}\right)^{1/2} \underset{q \simeq 1}\approx 0.08\,{\rm day}\,M_{\star,10}^{0.37},
\label{eq:Porb}
\ee  
is typically a few hours.  

As in the case of the original CE event itself, the mass-transfer process will be dynamically unstable for sufficiently high mass ratios $q \gtrsim q_{\rm dyn}$, where the precise threshold value $q_{\rm dyn}$ depends on the response of the donor to mass-loss and the how well the mass-transfer process conserves orbital angular momentum (e.g.,~\citealt{Ge+20}).  Early stripping of the He core during the CE phase is followed by a $\sim 100-1000$ yr phase of thermal readjustment \citep{VignaGomez+22}, which could impact this process.  A key feature destabilizing the mass-transfer process in this context is the high mass transfer rate $\sim M_{\star}/t_{\rm visc}^{\rm CD} \sim 10^{-3}-10^{-2}M_{\odot}$ yr$^{-1}$ driven by the torque from the outer circumbinary disk (for $t_{\rm visc}^{\rm CD} \sim 10^{3}-10^{4}$ yr; Eq.~\ref{eq:tvisc}), which exceeds by $\gtrsim 4$ orders of magnitude the BH/NS Eddington accretion rate and hence may evolved to dynamical instability, even under conservative assumptions (e.g., \citealt{King&Begelman99,Lu+22})

Dynamically unstable mass transfer leads to a runaway increase in the mass transfer rate (e.g., \citealt{MacLeod&Loeb20}, with implications for the CSM environment of the merger; Sec.~\ref{sec:CSM}) and, ultimately, the tidal disruption of the companion star \citep{Fryer&Woosley98,Zhang&Fryer01}.  This process concludes with the companion being quickly$-$in just a few orbits$-$sheared into an accretion disk surrounding the BH.  

\subsubsection{Disrupted WR Star Accretion Disk Properties}
\label{sec:accretion}

We now estimate the properties of the disk created from the disrupted WR star, immediately after its formation (an epoch we denote by the subscript `0').  The characteristic radial dimension of the disk can be estimated as (e.g., \citealt{Margalit&Metzger16}),
\be
R_{\rm d,0} \simeq a_{\rm RLOF}(1+q)^{-1} \underset{q = 1}\approx 1.3R_{\star} \approx 1.04R_{\odot}M_{\star,10}^{0.58}.
\label{eq:Rd0}
\ee
This is the semi-major axis of a point mass $\sim M_{\star}$ in orbit around the BH/NS, with angular momentum equal to that of the binary at the time of disruption (which is assumed to be conserved during the disruption process), and in the second equality we have used Eq.~\ref{eq:Rstar} and Eq.~\ref{eq:aRLOF} for $q = 1$.  

The mass of the formed disk will likewise approximately equal that of the original secondary, $M_{\rm d,0} \approx M_{\star}$, though a small fraction of the mass will fall promptly onto the surface of the compact object.  The initial surface density of the disk is then,
\be
\Sigma_{0} \approx \frac{M_{\star}}{2\pi R_{\rm d,0}^{2}} \approx 6\times 10^{11}\,{\rm g\,cm^{-2}}\,M_{\star,10}^{-0.16}.
\ee
Due to the gravitational energy released during the disruption process, and the inability to cool radiatively on the accretion timescale (see below), the disk will be hot and geometrically thick after forming, with a vertical scale-height $H_0$ and aspect ratio $\theta_0 \equiv H_0/R_{\rm d,0} \sim 1/3$ (e.g., \citealt{Metzger12}).  The characteristic midplane density at $r \sim R_{\rm d,0}$ is then given by,
\be
\rho_0 \approx \frac{\Sigma_{0}}{2H_0} \approx 14\,{\rm g\,cm^{-3}}\,M_{\star,10}^{-0.74}\theta_{0.33}^{-1},
\ee
where $\theta_{0.33} \equiv \theta_0/(0.33)$.

After forming, the disk will begin to accrete onto the BH/NS as a result of angular momentum transport driven by the MRI and gravitational instabilities \citep{Gammie01}.  In the case of MRI turbulence, the ``viscous'' timescale, over which the peak accretion rate is reached, is given by (e.g., \citealt{Frank+02})
\begin{eqnarray}
t_{\rm visc,0} &\sim& \left.\frac{r^{2}}{\nu}\right|_{R_{\rm d,0}} \sim \frac{1}{\alpha}\frac{1}{\theta_0^{2}}\left(\frac{R_{\rm d,0}^{3}}{GM_{\bullet}}\right)^{1/2} \nonumber \\
&\approx& 0.55\,{\rm day}\, \alpha_{0.1}^{-1}M_{\star,10}^{0.87}M_{\bullet,10}^{-0.5}\theta_{0.33}^{-2},  \label{eq:tvisc2}
\end{eqnarray}
where $\nu = \alpha c_{\rm s}H = \alpha r^{2}\Omega_{\rm K}\theta^{2}$ is again the effective kinematic viscosity and the other definitions are identical to those in Eq.~\ref{eq:tvisc}.
The viscous timescale is typically around 1 day, comparable to the peak timescale of FBOT light curves.

On timescales $t \gtrsim t_{\rm visc,0}$, the disk will establish a steady flow onto the BH/NS.  The peak accretion rate near the outer disk $\sim R_{\rm d,0}$,
\be
\dot{M}_0 \sim \frac{M_{\rm d,0}}{t_{\rm visc,0}} \sim 4\times 10^{29}{\rm g\,s^{-1}}\,\alpha_{0.1}M_{\star,10}^{0.13}M_{\bullet,10}^{0.5}\theta_{0.33}^{2},
\label{eq:Mdot0}
\ee
is $\gtrsim 10$ orders of magnitude larger than the Eddington rate $\dot{M}_{\rm Edd} \equiv L_{\rm Edd}/(0.1c^{2}) \sim M_{\bullet,10}10^{19}$ g s$^{-1}$, justifying our earlier assumption of a geometrically thick disk.

For mass ratios $q \sim 1$ of interest, the disk will be sufficiently massive to experience instabilities arising from self-gravity.  This occurs for values of the \citet{Toomre64} parameter,
\be
Q = \frac{\Omega c_s}{\pi G \Sigma} \approx \frac{\Omega_{\rm K}^{2}}{2\pi G \rho} \simeq \frac{M_{\bullet}(1+q)}{\pi r^{2}\Sigma}\theta \underset{r = R_{\rm d,0}}\sim \frac{1+q}{q}\theta,
\label{eq:Q}
\ee
less than a critical value $Q_0 \approx 2$.  The effect of such instabilities is to generate non-axisymmetric structures, such as spiral density waves, which mediate rapid angular momentum transport, quickly reducing the disk mass to the point of marginal stability $Q \approx Q_0$ (e.g., \citealt{Laughlin&Bodenheimer94,Gammie01}).  Because density waves can transport angular momentum over radial scales $\sim r$, the accretion timescale may be even shorter by a factor $\theta^{2} \sim 0.1$ than the value due to MRI turbulence alone (Eq.~\ref{eq:tvisc2}), i.e. hours instead of days.  

The practical effect of gravitational instabilities is therefore a prompt initial episode of accretion with a timescale as short as hours, until the disk mass is reduced sufficiently for $Q \gtrsim Q_0$ ($M_{\rm d}/M_{\bullet} \lesssim (Q_0/\theta_0-1)^{-1} \sim 0.3$) and MRI turbulence takes over.  This leads to reduced fraction of the disrupted star's mass being accreted on the MRI timescale $\sim t_{\rm visc,0} \sim$ 1 day (Eq.~\ref{eq:tvisc2}) and a peak accretion rate up to a few times lower than estimated for $\dot{M}_0$ in Eq.~\ref{eq:Mdot0}.

\subsubsection{Disk-Wind Outflows}
\label{sec:outflows}

For the high mass inflow rates $\dot{M} \gg \dot{M}_{\rm trap} \equiv \dot{M}_{\rm Edd}(R_{\rm d,0}/R_{\rm in}) \sim 10^{4} \dot{M}_{\rm Edd}$ of interest (see Eq.~\ref{eq:Mdot0}) photons are trapped and advected inwards through the disk at radii $\lesssim R_{\rm d,0}$ (e.g., \citealt{Begelman79}), where $R_{\rm in}$ corresponds to the inner edge of the disk (hereafter, we shall take $R_{\rm in} = 6GM_{\bullet}/c^{2} \sim 10^{7}M_{\bullet,10}$ cm corresponding to the innermost stable circular orbit of a non-spinning BH).  Since the disk cannot cool efficiently via radiation (e.g., \citealt{Shakura&Sunyaev73}), the accretion flow in this ``hyper-accretion'' regime is susceptible to outflows powered by the released gravitational energy (e.g., \citealt{Narayan&Yi95,Blandford&Begelman99,Kitaki+21}).  

Such outflows cause the mass inflow rate $\dot{M}$ to decrease approaching the BH/NS surface, in a way typically parametrized as a power-law in radius, 
\be
\dot{M}(r) \approx \dot{M}_0 \left(\frac{r}{R_{\rm d,0}}\right)^{p},
\label{eq:Mdotr}
\ee
where the the parameter $0 < p < 1$ \citep{Blandford&Begelman99} depends on details of the outflow model.  Throughout this paper we take $p = 0.6$ as fiducial, motivated by numerical simulations of radiatively inefficient accretion flows (e.g., \citealt{Yuan&Narayan14,Hu+22}).  

The prescription of Eq.~\ref{eq:Mdotr} predicts that large radii $\sim R_{\rm d,0} \sim 10^{11}$ cm in the disk dominate the total mass budget of the disk outflows, while the smallest radii $\sim R_{\rm in} \sim 10^{7}$ cm dominate their total energy budget.  The total mass-loss rate,
\be
\dot{M}_{\rm w} = \dot{M}(R_{\rm d,0})-\dot{M}(R_{\rm in}) = \dot{M}_{0}\left[1-\left(\frac{R_{\rm in}}{R_{\rm d,0}}\right)^{p}\right],
\label{eq:Mdotw}
\ee 
is almost equal to the total inflow rate $\sim \dot{M}_0$ because $R_{\rm d,0} \gg R_{\rm in}$.  Thus, most of the mass of the disrupted WR star will eventually be unbound, with only a small fraction reaching the BH/NS surface and accreting:
\begin{eqnarray}
\frac{M_{\rm acc}}{M_{\star}} &\approx& \frac{\dot{M}(R_{\rm in})}{\dot{M}(R_{\rm d,0})} \sim  \left(\frac{R_{\rm in}}{R_{\rm d,0}}\right)^{p} \sim 5\times 10^{-3}M_{\bullet,10}^{0.6}M_{\star,10}^{-0.35} \nonumber \\
&\Rightarrow& M_{\rm acc} \approx 0.05M_{\bullet,10}^{0.6}M_{\star,10}^{0.65}M_{\odot},
\label{eq:Macc}
\end{eqnarray}
where in the final numerical estimate we take $p = 0.6.$  This small quantity of mass accretion will not in general be sufficient to increase the mass of a NS accretor sufficiently to instigate collapse to a BH.

\citet{Margalit&Metzger16} estimate that the outflow speed of the disk winds is related to the local Keplerian orbital velocity $v_{\rm K} = r \Omega_{\rm K}$ according to $v_{\rm w} \approx 1.2 v_{\rm K}$, where the prefactor is again for $p = 0.6$ (\citealt{Margalit&Metzger16}, their Fig.~3).  Roughly half of the wind material will thus emerge from radii exterior to $R_{\rm w} \approx R_{\rm d,0}/2^{1/p} \sim 0.3R_{\rm d,0}$ with a mean velocity 
\begin{eqnarray}
v_{\rm slow} &\approx& 1.2 v_{\rm K}|_{R_{\rm w}} \approx 1.2\left(\frac{GM_{\rm bin}}{R_{\rm w}}\right)^{1/2} \nonumber \\
&&\underset{q =1}\approx 3900\,{\rm km\,s^{-1}}\,M_{\star,10}^{0.21},
\label{eq:vslow}
\end{eqnarray}
carrying a total kinetic energy,
\be
E_{\rm slow} \approx \frac{1}{2}M_{\star}v_{\rm slow}^{2} \approx 1.5\times 10^{51}{\rm erg}\,M_{\star,10}^{1.42}.
\label{eq:Eslow}
\ee
The velocity of this ``slow'' ejecta component are consistent with the lowest speeds $\sim 3000-4000$ km s$^{-1}$ measured for AT2018cow (e.g., \citealt{Perley+19,Margutti+19,Xiang+21}).

A smaller fraction of the disk-wind ejecta mass $\gtrsim M_{\rm acc}$ (Eq.~\ref{eq:Macc}) emerges from deeper in the gravitational potential well with much higher velocities $ \gg v_{\rm slow}$.  Radiation GRMHD simulations of super-Eddington accretion flows reveal the generation of trans-relativistic outflows from their innermost radii, carrying a combined radiative and kinetic luminosity $\sim L_{\rm acc} \sim \eta\dot{M}_{\bullet}c^{2}$ (e.g., \citealt{Sadowski&Narayan15,Sadowski&Narayan16}), where $\dot{M}_{\bullet} = \dot{M}(R_{\rm in})$ is the accretion rate reaching the BH/NS and $\eta$ is an efficiency factor.  Values of $\eta \sim 0.03$ were found by \citet{Sadowski&Narayan16} for accretion rates of a few hundred $\dot{M}_{\rm Edd}$, which are however still typically $\sim 4$ orders of magnitude smaller than the peak accretion rates under consideration here.  We shall nevertheless take $\eta \sim 10^{-2}$ as fiducial for the efficiency in what follows.  

Over the initial viscous timescale $t_{\rm visc,0} \sim$ 1 day after the merger (Eq.~\ref{eq:tvisc2}), such outflows will thus release a total energy in kinetic energy and radiation,
\be
E_{\rm fast} \sim \eta M_{\rm acc}c^{2} \sim 10^{51}{\rm erg}\,\eta_{-2}M_{\bullet,10}^{0.6}M_{\star,10}^{0.65},
\label{eq:Efast}
\ee
comparable or exceeding $E_{\rm slow}$ depending on the value of $\eta = 0.01\eta_{-2}$.  
As they interact with the wider-angle disk outflows, outflows from the inner disk will be collimated along the disk rotation axis (e.g., \citealt{DuPont+22}), generating a jet-like geometry and plausibly creating the fastest velocity $v_{\rm fast} \gtrsim 0.1$ c ejecta from LFBOTs (e.g., \citealt{Coppejans+20,Ho+20}), which carry a mass
\be
M_{\rm fast} = \frac{2E_{\rm fast}}{v_{\rm f}^{2}} \approx 0.1M_{\odot}\eta_{-2}M_{\bullet,10}^{0.6}M_{\star,10}^{0.65}\left(\frac{v_{\rm f}}{0.1c}\right)^{-2}.
\label{eq:Mfast}
\ee
The effective ``jet efficiency'' $f_{\rm mech} \equiv E_{\rm fast}/(0.1M_{\star}c^{2}) \sim 10^{-3}$ predicted here is similar to that adopted in previous works \citep{Schroder+20,Gottlieb+22}.

\subsubsection{Central Engine Luminosity}
\label{sec:engine}

At late times $t \gg t_{\rm visc,0}$, the outer edge of the disk will continue to spread outwards due to the redistribution of angular momentum.  If the disk outflows carry away only the local specific angular momentum of the disk material, then the outer edge of the disk will grow with time as (e.g., \citealt{Cannizzo+90})
\be
R_{\rm d} \simeq R_{\rm d,0}\left(\frac{t}{t_{\rm visc,0}}\right)^{2/3}, t \gg t_{\rm visc,0}.
\label{eq:Rd}
\ee
The accretion rate at radii $r < R_{\rm d}$ will likewise drop as a power-law in time (e.g., \citealt{Metzger+08}), viz.~
\be
\dot{M} \propto r^{p}t^{-4(p+1)/3}, t \gg t_{\rm visc,0},
\label{eq:Mdotlate}
\ee
Again taking $p = 0.6$, the accretion rate near the outer edge of the disk decays as,
\be
\dot{M}(R_{\rm d}) \sim \frac{\dot{M}_0}{3}\left(\frac{t}{t_{\rm visc,0}}\right)^{-1.47},
\ee
while the accretion rate reaching the BH/NS will decay more steeply in time,
\begin{eqnarray}
\dot{M}_{\bullet} \sim \frac{\dot{M}_{0}}{3}\left(\frac{R_{\rm in}}{R_{\rm d,0}}\right)^{p}\left(\frac{t}{t_{\rm visc,0}}\right)^{-\frac{4(p+1)}{3}}, t \gg t_{\rm visc,0}, \nonumber \\
\end{eqnarray}
where the additional prefactor of 1/3 accounts in a crude way for the reduction in disk mass due to rapid accretion from the early gravitationally unstable phase (Eq.~\ref{eq:Q} and related discussion).  This results in the fast-outflow luminosity (for $R_{\rm in} = 6 R_{\rm g}$ and $p = 0.6$),
\begin{eqnarray}
&&L_{\rm acc,0} \sim \eta \dot{M}_{\bullet}c^{2} \sim 1.6\times 10^{44}\,{\rm erg\,s^{-1}}\eta_{-2}\alpha_{0.1}^{-1.1} \times \nonumber \\
&&\theta_{0.33}^{-2.3}M_{\star,10}^{1.63}M_{\bullet,10}^{0.03}\left(\frac{t}{\rm 3\,day}\right)^{-2.1},\,\,\, t \gtrsim t_{\rm visc,0}, 
\label{eq:Lacc}
\end{eqnarray}
where we have used Eqs.~(\ref{eq:Rd0}), (\ref{eq:Mdot0}).  We observe that $L_{\rm acc}$ is broadly similar in normalization and power-law decay rate $\propto t^{-2.1}$ to the inferred engine luminosities of AT2018cow and other LFBOTs.  

The above formalism assumes that disk outflows carry away only the $\sim$Keplerian specific angular momentum of the disk.  However, the rate at which $\dot{M}_{\bullet}$ drops will be accelerated if the disk outflows carry away angular momentum more efficiently than this (e.g., by establishing an Alfv\'{e}n surface out of the disk midplane due to the presence of a strong ordered poloidal magnetic field; \citealt{Blandford&Payne81}).  In such cases, an exponential drop in the accretion power is expected on top of the canonical power-law decay $L_{\rm acc,0} \propto t^{-2.1}$ (Eq.~\ref{eq:Lacc}), of the form (\citealt{Metzger+08}, their Eq.~B15 for $p = 0.6)$
\be
L_{\rm acc} \approx L_{\rm acc,0}\exp\left[-\left(\frac{t}{t_{\rm visc,0}}\right)^{0.28}\right].
\label{eq:Lacc2}
\ee
Evidence for such a late-time steepening may be present in the X-ray light curves of AT2018cow \citep{Margutti+19} and AT2020xnd \citep{Bright+21,Perley+21,Ho+21} on timescales of $\sim$ 1 month $\sim 10 t_{\rm visc,0} \sim 10 t_{\rm pk}$ after the outburst.

On the other hand, as $L_{\rm acc}$ continues to decrease it will eventually begin to approach the Eddington luminosity $L_{\rm Edd} \approx 2\times 10^{39}M_{\bullet,10}$ erg s$^{-1}$ from above.  At this point the radiative efficiency of the accretion flow may increase from $\eta \lesssim 0.03$ (e.g., \citealt{Sadowski&Narayan16}) for super-Eddington accretion to the expected value $\eta \sim 0.1$ for a radiatively efficient BH accretion flows \citep{Novikov&Thorne73}.  This could lead to a temporary flattening of the X-ray light curve at very late times, e.g. $t \gtrsim 1$ yr.  In analogy to the state changes that occur in the X-ray binaries, the photon energy spectrum from the inner accretion flow would also soften during this transition, perhaps from something akin to the Comptonized ``ultra-luminous'' state observed in Ultraluminous X-ray sources \citep{Gladstone+09} to the soft thermal state \citep{Remillard&McClintock06}.   

Thermal emission from the outer accretion flow itself may also become visible at late times.  The ``trapping'' radius, exterior to which photons escape the disk on the inflow time, is given by $R_{\rm trap} \simeq (R_{\rm in}/2)(10\dot{M}_{\bullet}/\dot{M}_{\rm Edd})$ \citep{Begelman79}.  Using Eqs.~\ref{eq:Mdotlate}, \ref{eq:Lacc}, we find:
\begin{eqnarray}
R_{\rm tr} &\approx& R_{\rm in}\left(\frac{L_{\rm acc}}{2\eta L_{\rm Edd}}\right)^{\frac{1}{1-p}} \underset{p = 0.6}\approx \nonumber \\
&& 0.3R_{\odot}\, \alpha_{0.1}^{-2.8}\theta_{0.33}^{-5.8}M_{\star,10}^{4.1}M_{\bullet,10}^{-1.4}\left(\frac{t}{1000\,{\rm days}}\right)^{-5.3},
\label{eq:Rtr}
\end{eqnarray}
resulting in values $R_{\rm tr} \sim 0.1-10R_{\odot}$ depending on the parameters.  The luminosity of the disk from radii $\lesssim R_{\rm tr}$ is approximately equal to the Eddington luminosity, $L_{\rm tr} \sim L_{\rm Edd} \approx 2\times 10^{39}M_{\bullet,10}$ erg s$^{-1}$.  The blackbody radius of the emission is
\be
T_{\rm tr} \simeq \left(\frac{L_{\rm tr}}{4\pi \sigma R_{\rm tr}^{2}}\right)^{1/4} \approx 5\times 10^{4}\,{\rm K}\,M_{\bullet,10}^{1/4}\left(\frac{R_{\rm ph}}{10R_{\odot}}\right)^{-1/2},
\label{eq:Ttr}
\ee
where $R_{\rm ph} \gtrsim R_{\rm tr}$ is the photosphere radius.  

It is tempting to associate this emission with the thermal optical/UV source detected at $t \sim 1000$ days from AT2018cow \citep{Sun+22} with $L 
\gtrsim 10^{40}$ erg s$^{-1}$; $T_{\rm eff} \gtrsim 4\times 10^{4}$ K for a massive $\sim 100M_{\odot}$ BH accretor.  However, the temporal stability of the optical/UV flux (corresponding to change of $\lesssim 0.1$ mag between two epochs separated by over a year) from what should be an evolving accretion flow, poses a challenge to this scenario.

\subsubsection{Composition of the Disk-Wind Ejecta}
\label{sec:burning}

The bulk of the disk-wind ejecta will occur from radii $\sim R_{\rm d,0}/3 \sim 1R_{\odot}$ and is composed of unprocessed material from the disrupted WR star (e.g., trace H, $^{4}$He, $^{12}$C, $^{14}$N, $^{16}$O), depending on its state of nuclear evolution (the WN/WC/WO classification of isolated WR stars; e.g., \citealt{Crowther07}).  However, heavier elements can be synthesized at smaller radii in the disk $\lesssim 10^{9}$ cm and then carried outwards with the disk-wind ejecta.  In particular, we expect nuclear burning to occur in the disk midplane qualitatively similar to that in collapsars (e.g., \citealt{MacFadyen&Woosley99,Zenati+20}) and the merger of a white dwarf with a NS or BH (e.g., \citealt{Metzger12,Fernandez&Metzger13,Zenati19b,Fernandez+19,Zenati+20,Bobrick+21}).  Intermediate radii in the disk $\lesssim 10^{8.5}$ cm are hot enough to burn lighter elements, while heavier elements require the greater temperatures at smaller radii $\lesssim 10^{7.5}$ cm \citep{Metzger12}.  This chain can in principle extend all the way up to the Fe group, with the production of $^{56}$Ni occurring in a narrow annular region at temperatures $T_{\rm Ni} \sim 4\times 10^{9}$ K.  

At small radii in the disk $r \ll R_{\rm d,0}$ radiation pressure dominates gas pressure.  The radial temperature profile on timescales $\sim t_{\rm visc,0} \sim $ days (when most of the mass-loss occurs) is given by:
\begin{eqnarray}
&&T(r,t_{\rm visc,0}) \simeq \left[\frac{3GM_{\bullet}\rho}{a r}\theta^{2}\right]^{1/4} \nonumber \\
&&\simeq 2.6\times 10^{9}{\rm K}\,\theta_{0.33}^{0.5}M_{\bullet,10}^{-0.45}M_{\star,10}^{-0.11}\left(\frac{r}{6R_{\rm g}}\right)^{-0.48},
\label{eq:T0}
\end{eqnarray}
where $\rho(r,t_{\rm visc,0}) \simeq \dot{M}_{\bullet}(t_{\rm visc,0})/(6\pi \alpha r^{2}v_{\rm K}\theta^{3})$ is the steady-state midplane density profile.

This illustrates that the synthesis of $^{56}$Ni at $T \approx T_{\rm Ni}$ is possible only very close to the BH, at radii $r \lesssim 6R_{\rm g}$, resulting in at most $M_{\rm Ni} \sim M_{\rm acc} \sim few \times 10^{-2}M_{\odot}$ (Eq.~\ref{eq:Macc}).  Detailed multidimensional simulations of white dwarf/NS mergers find even smaller values $M_{\rm Ni} \lesssim 10^{-3}-10^{-2}M_{\odot}$ \citep{Fernandez+19}, though somewhat larger $^{56}$Ni yields may be possible if nuclear detonations occur within the disk \citep{Zenati+20}.  Such small $^{56}$Ni abundances are consistent with light curve modeling of LFBOT and Type Icn SNe (e.g., \citealt{Perley+19}).

Intermediate mass-elements, synthesized in greater abundance in the disk, include $\alpha$-capture elements such as $^{20}$Ne, $^{24}$Mg, $^{28}$Si, $^{40}$Ca, $^{52}$Fe.  Given the He-rich nature of the disrupted WR star (similar to a He or mixed He/C/O composition white dwarf), previous studies show the disk will be particularly efficient at generating Ne and Ca (see Table D1 of \citealt{Margalit&Metzger16}), which are observed to be present in the ejecta of Type Icn SNe \citep{Gal-Yam+22}.  A more detailed calculation of the disk nucleosynthesis in the context of WR-BH/NS merger disks is required to better quantify the disk-wind ejecta.

\section{Transient Emission}
\label{sec:transient}

We now describe the sources of electromagnetic emission which accompany the merger-initiated, WR star-accretion-powered explosion described above.  We begin in Section \ref{sec:CSM} with an overview of the various sources of CSM surrounding the binary at the time of the merger, most of which are concentrated in the binary equatorial plane.  Then in Section \ref{sec:lightcurves} we outline a toy model for the X-ray and optical emission, the latter powered by a combination of reprocessing of X-rays from the accretion-powered jet by the fast polar outflow and shock emission between the slower bulk ejecta and equatorial CSM.  In Section \ref{sec:radio} we describe how the radio and mm emission is generated on larger scales by shock interaction between the fast component and the outermost CSM.  These different sources of emission are summarized in the bottom panel of Fig.~\ref{fig:cartoon}.

\subsection{Sources of Gaseous Circumbinary Medium}
\label{sec:CSM}

The multi-component disk-wind ejecta described in the previous section can possess a large kinetic energy, $E_{\rm slow}, E_{\rm fast} \gtrsim 10^{51}$ erg (Eqs.~\ref{eq:Eslow}, \ref{eq:Efast}), but much of this would not be observed as radiation if the outflows were to expand into vacuum.  The disk-wind ejecta is extremely opaque near the launching radii $\lesssim R_{\rm d,0} \sim R_{\odot}$ (for the same reasons that the accretion flow itself is radiatively inefficient) and so will lose most of its internal energy to $PdV$ expansion prior to reaching large enough radii (low enough optical depth) to radiate.

In actuality, the disk outflows are unlikely to emerge into a low density environment.  Various sources of dense gas will surround the binary at the time of the merger/tidal disruption, whose interaction with the faster merger ejecta will generate a powerful electromagnetic signal.  We summarize these sources of CSM here, starting at small radii closest to the merging binary and moving outwards.

\subsubsection{Pre-Merger WR Mass-Loss from $L_2$}  The merger is not an instantaneous process.  One source of circumbinary CSM will arise due to mass-loss from the WR star following the onset of unstable mass-transfer but prior to the final tidal disruption \citep{Pejcha+16a,Pejcha+16b,MacLeod+17,Pejcha+17,MacLeod&Loeb20}.  This mass-loss can occur from the outer $L_2$ Lagrange point in the form of a wide-angle outflow concentrated in the binary plane (e.g., \citealt{Pejcha+16a,Pejcha+17,Lu+22}).  Such material can possess a mass up to $\sim 15\%$ that of the donor star, i.e. $M_{\rm L2} \approx 0.15M_{\star} \sim M_{\odot}$ (e.g., \citealt{MacLeod+17}) and achieve a velocity comparable to the binary escape speed \citep{Pejcha+16b}, 
\be
v_{\rm esc} \simeq \left(\frac{G M_{\rm bin}}{a_{\rm RLOF}}\right)^{1/2} \underset{q=1}\approx 1400\,{\rm km\,s^{-1}}\,M_{\star,10}^{0.21},
\label{eq:vL2}
\ee
though the outflow velocity at the beginning of the runaway may start several times slower than this \citep{Pejcha+16a}.  

If the runaway phase of unstable mass-loss begins $N \sim 10-100$ orbital periods prior to the dynamical phase (i.e., a time $t_{\rm run} \sim NP_{\rm orb} \sim 8\,(N/100)M_{\star,10}^{0.37}$ {\rm days}\, using Eq.~\ref{eq:Porb}; \citealt{MacLeod&Loeb20}), then by the time of the final merger this material could reach radii
\be
R_{\rm L2} \sim v_{\rm esc}t_{\rm run} \approx 9\times 10^{13}{\rm cm}\,\left(\frac{N}{100}\right)M_{\star,10}^{0.58}.
\label{eq:RL2}
\ee
Detailed modeling of the well-studied low-mass stellar merger event V1309 Sco \citep{Tylenda+11} implied $N > 5-20$ \citep{Pejcha14,Pejcha+17}.  Indirect evidence based on modeling the light curves of more massive stellar merger events supports the presence of dense circumbinary gas on radial scales $\sim 10^{14}-10^{15}$ cm \citep{Matsumoto&Metzger22}. 

\subsubsection{Relic Circumbinary Disk from CE Phase} Another source of circumbinary material is the relic bound disk from the first mass-transfer phase or CE event (e.g., \citealt{Kashi&Soker11}), the torques from which may be necessary to drive the binary together in the first place (Sec.~\ref{sec:delayed}).  At the time of the merger, the disk could possess a mass up to $M_{\rm CD} \sim 0.01-0.1M_{\rm bin} \sim 0.1-1M_{\odot}$ concentrated at radii,
\begin{eqnarray}
R_{\rm CD} &\sim& a_0\left(\frac{M_{\rm bin}}{M_{\rm CD}}\right)^{2} \nonumber \\
&\approx& 6\times 10^{14}{\rm cm}\left(\frac{a_0}{100R_{\odot}}\right)\left(\frac{M_{\rm CD}}{0.1M_{\rm bin}}\right)^{-2}. 
\label{eq:RCD}
\end{eqnarray}
This predicts $R_{\rm CD} \sim 10^{14}-10^{16}$ cm for values of the post-CE binary semi-major axis $a_0 \gtrsim 30R_{\odot}$ and circumbinary disk mass $M_{\rm CD}\lesssim 0.1M_{\rm CD}$ which give long enough WR-BH/NS merger times ($t_{\rm visc}^{\rm CD} \gg 10^{3}$ yr; Eq.~\ref{eq:tvisc}) to clear out most of the unbound H-rich envelope from the first mass-transfer/CE event (Sec.~\ref{sec:delayed}).  

\subsubsection{Circumbinary Disk Outflows} 
\label{sec:binarydiskoutflows}
Winds from the circumbinary disk can generate a source of CSM extending to yet larger radii $\gg R_{\rm CD}$.    One efficient mass-loss mechanism is photoevaporation by the WR star due to its high ionizing luminosity $L_{\star} \sim L_{\rm Edd} \sim 2\times 10^{39}(M_{\star}/10M_{\odot}){\rm erg\,s^{-1}}$, similar to what creates hyper-compact HII regions around massive proto-stars (e.g., \citealt{Keto07}).  Mass-loss due to photoevaporation becomes important external to the critical radius,
\be
R_{\rm g} \simeq \frac{2GM_{\rm bin}}{c_{\rm s}^{2}} \sim 10^{16}{\rm cm}\left(\frac{M_{\rm bin}}{30M_{\odot}}\right),
\label{eq:Rg}
\ee
at which the escape speed of the disk equals the sound speed $c_{\rm s} \approx 10$ km s$^{-1}$ of $\sim 10^{4}$ K photoionized gas \citep{Hollenbach+94}.

The rate of mass-loss from the circumbinary disk at radii $\sim R_{\rm g}$ is approximately given by  \citep{Hollenbach+94}
\be
  \dot{M}_{\rm ph} \approx 7\times 10^{-5}M_{\odot}\,{\rm yr^{-1}}\left(\frac{\Phi_i}{10^{50}{\rm s^{-1}}}\right)^{1/2}\left(\frac{M_{\rm bin}}{30M_{\odot}}\right)^{1/2},
\label{eq:Mdotph}
\ee
where $\Phi_i \sim L_{\star}/\epsilon_{\rm H} \sim 10^{50}(M_{\star}/10M_{\odot})\,{\rm s^{-1}}$ is the WR star's Lyman continuum photon luminosity and $\epsilon_{\rm H} = 13.6$ eV.  At this rate of mass-loss, a circumbinary disk of mass $M_{\rm CD} \sim 0.01-0.1M_{\rm bin} \sim 0.1-1M_{\odot}$ would completely evaporate in around $\sim 10^{3}-10^{4}$ years, shorter than the WR lifetime. 

Given the velocity of such winds from the circumbinary disk, $v_{\rm w} \sim c_{\rm s} \sim 10$ km s$^{-1}$, the implied gas density on scales $\sim R_{\rm g}$ is given by
\begin{eqnarray}
n_{\rm ph} &\sim& \frac{\dot{M}_{\rm ph}}{4\pi v_{\rm w}r^{2}m_p} \sim 10^{6}\,{\rm cm^{-3}}\left(\frac{M_{\star}}{10M_{\odot}}\right)^{1/2}\times \nonumber \\
&& \left(\frac{M_{\rm bin}}{30M_{\odot}}\right)^{1/2}\left(\frac{r}{10^{16}{\rm cm}}\right)^{-2}, 
\label{eq:nph}
\end{eqnarray}
comparable to the environments on radial scales $\gtrsim 10^{16}$ cm surrounding LFBOTs as inferred from their radio/mm emission (Sec.~\ref{sec:radio}).  Other forms of disk mass-loss could also occur, for instance stripping by the WR star wind (e.g., \citealt{Elmegreen78}), can also contribute to the disk removal process (e.g., \citealt{Hollenbach+00}).

The first CSM source (runaway mass-loss from the WR leading up to the merger) will be present and similar in its properties regardless of the larger-scale environment surrounding the WR-BH/NS merger.  However, the presence, quantity and radial scale of the second two CSM sources will depend sensitively on the properties of the bound debris disk from the original mass-transfer or CE event and the delay until the merger after the disk's formation.  This is important to the argument that LFBOTs and some Type Ibn/Icn SNe could share a related origin (Sec.~\ref{sec:unification}).  Also note that the first two CSM sources will interact with each other prior to the merger.  The WR ejecta from the immediate pre-merger phase will collide with the (effectively stationary) relic CE disk, which will act to decelerate the former and may could give rise to a phase of shock-powered precursor emission lasting days prior to the main merger-powered transient.

\subsection{Optical and X-ray Light Curve Model}
\label{sec:lightcurves}

The optical light curves during the first few weeks after the merger can be powered by two sources: (1) shock interaction between the disk-wind ejecta of the disrupted WR star and the CSM surrounding the original binary on radial scales $\lesssim 10^{15}$ cm; and (2) reprocessing of X-rays from the inner accretion flow or jet by the faster polar outflow; the observed X-ray emission in this scenario is just the fraction that avoid reprocessing.   

\subsubsection{Shock-Powered Emission}
\label{sec:shock}

We first consider emission powered by the collision between the slow disk-wind ejecta (mass $M_{\rm slow} \sim M_{\star}$; velocity $v_{\rm slow} \sim 3000$ km s$^{-1}$; Eq.~\ref{eq:vslow}) and the even slower pre-merger CSM (mass $M_{\rm pre} \sim 0.1M_{\star}$; velocity $v_{\rm pre} \lesssim v_{\rm esc} \lesssim v_{\rm slow}/2;$ Eq.~\ref{eq:vL2}).  We focus on the CSM arising from the WR RLOF mass-loss leading up to the merger (the first CSM source described in Sec.~\ref{sec:CSM}) instead of relic CE disk material.  As just described, the former is typically more massive/faster and thus will overtake the latter before the arrival of the even faster post-merger disk-wind ejecta.

We approximate the WR outflow leading up to the merger as that of an equatorially-focused wind (e.g., from the $L_{2}$ point) of constant velocity $v_{\rm pre}$, with a radial density profile
\be
\rho_{\rm pre}(r) = \frac{\dot{M}_{\rm pre}(t_{\rm pre} = r/v_{\rm pre})}{4\pi f_{\Omega} v_{\rm pre}r^{2}},
\label{eq:rhopre}
\ee
where $\dot{M}_{\rm pre}$ is the wind mass-loss rate at time $t_{\rm pre}$ prior to the tidal disruption/dynamical merger, $f_{\Omega} \approx 0.3$ is the fraction of the total solid angle subtended by the wind (e.g., \citealt{Pejcha+16a}).  Following \citet{Metzger&Pejcha17}, we parameterize the runaway mass-loss rate from the WR star leading up to the merger time as an exponential,
\be
\dot{M}_{\rm pre}(t_{\rm pre}) = \frac{M_{\rm pre}}{t_{\rm run}}\exp\left(-\frac{t_{\rm pre}}{t_{\rm run}}\right),\,\,\, t_{\rm pre} > 0,
\ee
where $t_{\rm run} = Nt_{\rm orb}$ and $N \sim 10-100$ is the number of binary orbital periods over which the mass-loss grows approaching the merger.  


At the onset of the WR-BH/NS merger driven-explosion ($t > 0$), the disk-wind ejecta will drive a shock into the slower pre-dynamical wind in the binary plane, thus acting to slightly decelerate the faster disk-wind ejecta and convert a fraction of its kinetic energy into radiation.  At the same time, in the comparatively empty polar regions, the disk-wind ejecta will quickly pass (and ``wrap around'') the equatorial ejecta shell (e.g., \citealt{Metzger10,Andrews&Smith18}).  

Again following previous work (e.g., \citealt{Metzger&Pejcha17}), we evolve the thermal energy $\mathcal{E}$ of radiation in the equatorial region of the disk-wind ejecta in a one-zone approximation according to:
\be
\frac{d\mathcal{E}}{dt} = -\frac{\mathcal{E}}{t} - L_{\rm opt,sh} + L_{\rm sh},
\label{eq:dEdt}
\ee
where the first term accounts for $PdV$ losses and the second term for radiative losses, where
\be
L_{\rm opt,sh} = \frac{\mathcal{E}}{t_{\rm diff} + t_{\rm lc,slow}} 
\label{eq:Loptsh}
\ee
is the radiated optical luminosity,
\be
t_{\rm diff} = \frac{M_{\rm slow}\kappa_{\rm slow}}{4\pi R_{\rm slow}c}
\ee
is the photon diffusion time through the slow ejecta of radius $R_{\rm slow} = v_{\rm slow}t$, and $t_{\rm lc,slow} = R_{\rm slow}/c$ is the light-crossing time.  For simplicity, we assume a constant opacity $\kappa_{\rm slow} \approx 0.03$ cm$^{2}$ g$^{-1}$ in the range expected for H-depleted ejecta (e.g., \citealt{Kleiser&Kasen14}).

The final term in Eq.~(\ref{eq:dEdt}) accounts for the luminosity due to shock heating, which we calculate assuming momentum-conserving interaction and $M_{\rm pre} \ll M_{\rm slow}$, according to
\begin{eqnarray}
L_{\rm sh} &\approx& 4\pi f_{\Omega}R_{\rm sh}^{2}\rho_{\rm pre}(R_{\rm sh})(v_{\rm sh}^{2}/2)v_{\rm sh} \nonumber \\
&\approx& \frac{1}{2}\frac{M_{\rm pre}v_{\rm slow}^{3}}{v_{\rm esc} t_{\rm run}}\exp\left[-\frac{v_{\rm slow}t}{v_{\rm esc}t_{\rm run}}\right]
\end{eqnarray}
where $R_{\rm sh} \simeq R_{\rm slow} = v_{\rm slow}t$ is the shock radius and we take $v_{\rm pre} = v_{\rm esc}$.  In assuming the forward shock expands at the same velocity as the slow ejecta ($v_{\rm sh} \simeq v_{\rm slow}$) we are neglecting deceleration of the ejecta, as justified to leading order if $M_{\rm pre} \ll M_{\rm slow}$. 

\subsubsection{X-ray Reprocessing}
\label{sec:Xray}

The $\sim 1000$ km s$^{-1}$ radiative shocks described above generate X-rays, but the high densities of the surrounding gas from the slow  outflow and CSM guarantee that the shock's X-ray luminosity is efficiently absorbed and thermalized into optical radiation, except possibly at very late times.  By contrast, for X-rays released along the polar axis from the inner accretion flow (e.g., \citealt{Pasham+21}), or by the interaction of a relativistic jet with the surrounding wind medium (e.g., \citealt{Gottlieb+22}), a greater fraction will escape because of the much lower density of the fast polar outflow.  A second component of optical emission that we consider is the partial reprocessing of the central X-ray source of luminosity $L_{\rm acc}$ (Eq.~\ref{eq:Lacc}) by the fast polar ejecta of mass $M_{\rm fast} \sim 0.1M_{\odot}$ (Eq.~\ref{eq:Mfast}) and velocity $v_{\rm fast} \sim 0.1$ c.

Analogous to Eq.~(\ref{eq:dEdt}) for the equatorial ejecta, we follow the time-evolution of the thermal energy $\mathcal{E}_{\rm pol}$ of radiation contained in the fast polar ejecta, also in a one-zone approximation, according to:
\be
\frac{d\mathcal{E}_{\rm pol}}{dt} = -\frac{\mathcal{E}_{\rm pol}}{t} - L_{\rm opt,rep}  + L_{\rm acc,th},
\label{eq:dEpoldt}
\ee
where again the first term accounts for $PdV$ losses, the second term for optical radiative losses, 
\be
L_{\rm opt,rep} = \frac{\mathcal{E}_{\rm pol}}{t_{\rm diff} + t_{\rm lc,fast}},
\label{eq:Loptrep}
\ee
where $t_{\rm lc,fast} = R_{\rm fast}/c$ and now
\be
t_{\rm diff} = \frac{M_{\rm fast}\kappa_{\rm fast} }{4\pi R_{\rm fast} c}
\ee
is the radial diffusion time through the fast ejecta of radius $R_{\rm fast} \simeq v_{\rm fast}t$, where the opacity $\kappa_{\rm fast} \approx \kappa_{\rm es} \approx 0.2$ cm$^{2}$ g$^{-1}$ corresponds to electron scattering for fully-ionized H-depleted gas (the polar outflow is photoionized by the luminous central X-ray source; \citealt{Margutti+19}).  

The final term in Eq.~\ref{eq:dEpoldt} is the fractional amount of the intrinsic accretion-powered X-ray luminosity, $L_{\rm acc} \propto t^{-2.1}$ (Eqs.~\ref{eq:Lacc}, \ref{eq:Lacc2}), which is thermalized into optical radiation according to
\be
L_{\rm acc,th} = \phi_0\left(1-e^{-\tau_{\rm X}}\right)L_{\rm acc} + \phi_0 L_{\rm acc},
\label{eq:Laccth}
\ee
where
\be
\tau_{\rm X} = \frac{M_{\rm fast}\kappa_{\rm X}}{4\pi R_{\rm fast}^{3}}
\ee
is the optical depth for X-ray thermalization and $\kappa_{\rm X}$ is the effective opacity for absorbing and thermalizing X-rays.  For $L_{\rm acc}$ we use Eq.~(\ref{eq:Lacc2}) at times $t > t_{\rm visc,0}$ (Eq.~\ref{eq:tvisc2}) and take $L_{\rm acc} = L_{\rm acc}(t = t_{\rm visc,0})$ at times $t < t_{\rm visc,0}.$  The escaping X-ray luminosity is likewise given by
\be
L_{\rm X} = L_{\rm acc} - L_{\rm acc,th} = \phi_0 L_{\rm acc}e^{-\tau_{\rm X}}.
\label{eq:Lx}
\ee
The factor of $\phi_0$ in Eqs.~\ref{eq:Laccth}, \ref{eq:Lx} is the assumed fixed fraction (taken fiducially to be $\phi_0 = 1/2$) of the solid angle subtended by the slow outflow, which will absorb and reprocess X-rays from the central engine even after the fast polar outflow is optically thin.   

A detailed model for the ionization state of the polar ejecta, beyond the scope of this work, is required to accurately determine $\kappa_{\rm X}$; in what follows, we take $\kappa_{\rm X} \sim 2\kappa_{\rm es} = 0.4$ cm$^{2}$ g$^{-1}$, i.e. $\tau_{\rm X} \sim 2\tau_{\rm es}$.  This choice is motivated by the fact that the probability of thermalizing an X-ray increases rapidly with the number of scatterings it experiences $N_{\rm scatt} \propto \tau_{\rm es}^{2}$, which is a rapidly increasing function of the Thomson optical depth 
\be
\tau_{\rm es} \approx \frac{M_{\rm fast}\kappa_{\rm fast}}{4\pi(v_{\rm fast}t)^{2}} \sim 2\left(\frac{M_{\rm fast}}{0.1M_{\odot}}\right)\left(\frac{v_{\rm fast}}{0.1c}\right)^{-2}\left(\frac{t}{\rm 10\,day}\right)^{-2}.
\ee 

In both the fast and slow ejecta components, we neglect heating due to the radioactive decay of $^{56}$Ni and $^{56}$Co.  Given the low-synthesized Ni mass $\lesssim 10^{-2}M_{\odot}$ in the disk outflows (Sec.~\ref{sec:burning}) their expected contribution to the light curve will typically be small compared to those from shocks or X-ray reprocessing.

\subsubsection{Example Light Curves: Application to AT2018cow}
\label{sec:example}

\begin{figure*}
    \centering
    \includegraphics[width=1.0\textwidth]{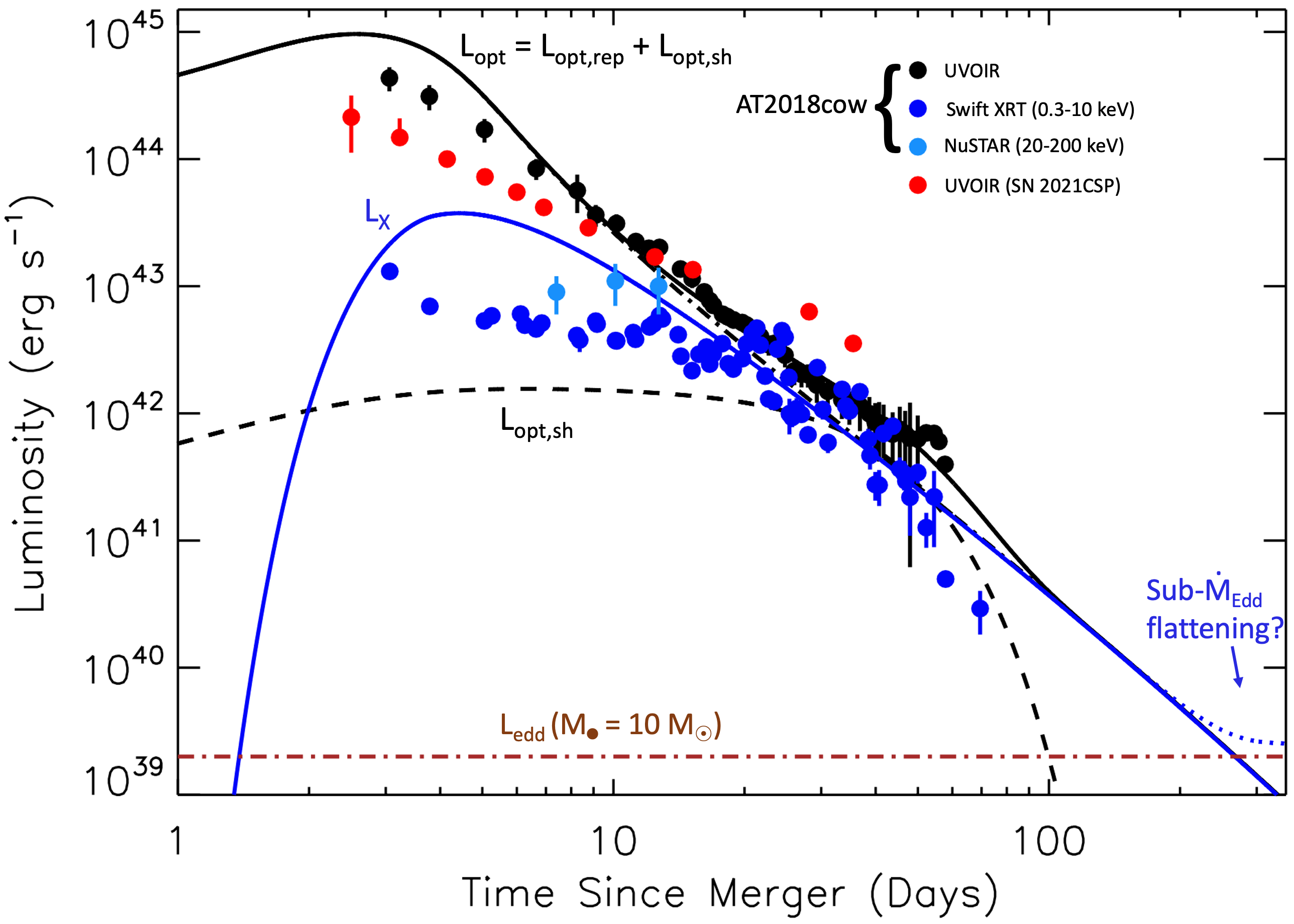}
    \caption{Example optical/X-ray light curve model confronts data.  The total optical luminosity $L_{\rm opt} = L_{\rm acc,rep} + L_{\rm opt,sh}$ (black solid line) includes reprocessed energy from the central engine $L_{\rm opt,rep}$ (black dot-dashed line) and CSM shock interaction $L_{\rm opt,sh}$ (black dashed line).  The escaping X-ray luminosity $L_{\rm X}$ is shown as a solid blue line.  We have assumed the merger of an equal-mass binary $M_{\star} = M_{\bullet} = 10M_{\odot}$, giving rise to an accretion disk with viscosity $\alpha = 0.03$ and slow/fast disk-wind ejecta with the following properties: $M_{\rm slow} = 9.8M_{\odot}$, $v_{\rm slow} = 3000$ km s$^{-1}$, $\kappa_{\rm slow} = 0.03$ cm$^{2}$, $M_{\rm fast} = 0.1M_{\odot}$, $v_{\rm fast} = 0.2c$, $\kappa_{\rm fast} = 0.2$ cm$^{2}$ (all close to fiducial values motivated in Sec.~\ref{sec:outflows}, \ref{sec:lightcurves}).  For the CSM, we assume $L_2$ mass-loss from the WR of mass $M_{\rm pre} = 0.1M_{\star} = 1M_{\odot}$ released $N = 30$ orbits prior to the merger (Eq.~\ref{eq:rhopre}).  For the central X-ray source we assume a luminosity $L_{\rm acc} = \eta \dot{M}_{\bullet}c^{2}$ with $\eta = 0.03$ and an effective reprocessing opacity $\kappa_{\rm X} = 2\kappa_{\rm fast}$.  Shown for comparison with circles are light curve observations of AT2018cow at optical (``bolometric'' UVOIR; black circles), {\it Swift} XRT soft X-ray (0.3-10 keV; dark blue circles) and {\it NuSTAR} hard X-ray (20-200 keV; light blue circles) energies from \citet{Margutti+19}.  Also shown is the UVOIR optical light curve for the Type Icn SN 2021csp (red circles; \citealt{Perley+22}).  The dotted blue line illustrates schematically a flattening of the X-ray light curve that may occur as the BH accretion rate approaches the Eddington value (luminosity $L_{\rm Edd} \approx 2\times 10^{39}$ erg s$^{-1}$; brown dot-dashed line).  We note that although the optical light curve is expected to be relatively isotropic, the X-ray light curve may exhibit a significant dependence on the inclination angle relative to the fast outflow axis, which could boost or reduce the isotropic-equivalent luminosity relative to the total luminosity estimated here.} 
    \label{fig:example}
\end{figure*}

Figure \ref{fig:example} shows an example of the X-ray and optical light curves, calculated by solving Eqs.~\ref{eq:dEdt},\ref{eq:dEpoldt} for the canonical case of an equal mass binary $M_{\star} = M_{\bullet} = 10M_{\odot}$ with their associated slow equatorial $\{M_{\rm slow} = 9.8M_{\odot}$, $v_{\rm slow} = 3000$ km s$^{-1}$\} and fast polar $\{M_{\rm fast} = 0.1M_{\odot} \sim M_{\rm acc}$, $v_{\rm fast} = 0.2$ c$\}$ ejecta components following their analytically expected values (Sec.~\ref{sec:outflows}), which roughly match those observed or inferred for AT2018cow.  For the optical light curve, we consider separately emission powered by shock interaction with the immediate pre-merger WR mass-loss $L_{\rm opt,sh}$ (Eq.~\ref{eq:Loptsh}; for characteristic CSM properties $M_{\rm pre} = 0.1M_{\star} = 1 M_{\odot}$, $v_{\rm pre} = v_{\rm slow}/2$; $N = 30$) and that due to reprocessing of the engine's X-rays $L_{\rm opt,rep}$ (Eq.~\ref{eq:Loptrep}; calculated for $\kappa_{\rm X} = 2\kappa_{\rm fast}$ and engine properties $\eta = 0.03$, $\alpha = 0.03$), as shown with dashed and dot-dashed lines, respectively.  Conversely, the escaping X-ray luminosity from the inner accretion flow, $L_{\rm X}$ (Eq.~\ref{eq:Lx}), is shown as a solid blue line.

Shown for comparison are the ``bolometric" UVOIR light curves of AT2018cow (black dots; \citealt{Margutti+19}) and the Type Icn SN 2021CSP (red dots; \citealt{Perley+22}), as well as the soft X-ray (0.3-10 keV) and hard X-ray (10-200 keV) light curves of AT2018cow (dark and light blue dots; \citealt{Margutti+19}).  Although our light curve model is highly simplified$-$for example, treating what is undoubtedly a complex jet/ejecta/CSM angular structure with a 2-zone treatment$-$it nevertheless reproduces many of the qualitative features of LFBOTs.  

The optical luminosity during the first week is dominated by reprocessing by the fast ejecta by X-rays from the central accretion source, with $L_{\rm opt} \propto L_{\rm acc} \propto t^{-2.1}$ (Eq.~\ref{eq:Lacc}).  By contrast, the luminosity from CSM shock interaction becomes of comparable importance around $t \sim 20$ days, around when the optical spectral features in AT2018cow showed an abrupt transition from broad spectral features to substantially narrower spectral features (e.g., \citealt{Perley+19,Margutti+19}) with asymmetric shape indicative of reprocessing by a torus-shaped medium \citep{Margutti+19}.  

At early times, the X-ray luminosity is suppressed relative to the optical because most of the X-rays are being absorbed and reprocessed by the fast ejecta shell ($\tau_{\rm X} \gg 1$); however, by $t \sim 10$ days when $\tau_{\rm X} \sim \tau_{\rm es} \sim 1$ a large fraction of the X-rays escape without absorption and $L_{\rm X} \sim L_{\rm opt}.$  The same Thomson depth $\tau_{\rm es} \sim few$ through the fast shell is consistent with the observation and then fading of the Compton hump feature detected by NuSTAR on this timescale \citep{Margutti+19}.  
The overall normalization of the optical/X-ray light curves reasonably matches that of AT2018cow for the assumed value $\eta \sim 0.03$ for the X-ray efficiency of the polar accretion flow, similar to those found by GRMHD simulations of super-Eddington accretion flows (e.g., \citealt{Sadowski&Narayan16}).  

The steepening of the X-ray light curve starting around 1 month is not reproduced by our model, but could in principle result from a reduced BH accretion rate due to either (1) non-conservation of specific angular momentum in the disk outflows (e.g., \citealt{Metzger+08}; see Eq.~\ref{eq:Lacc2}); (2) expansion of the outer edge of the growing disk (Eq.~\ref{eq:Rd}) beyond the photon-trapping radius \citep{Begelman79}, resulting in a thinner disk and longer viscous timescale.  On the other hand, as described in Sec.~\ref{sec:engine}, the X-ray light curve may (temporarily) flatten at even later times, once $L_{\rm X}$ approaches the BH Eddington luminosity $\approx 2\times 10^{39}$ erg s$^{-1}$; we denote this flattening schematically with a dotted blue line in Fig.~\ref{fig:example}.

\begin{figure}
    \centering
    \includegraphics[width=0.5\textwidth]{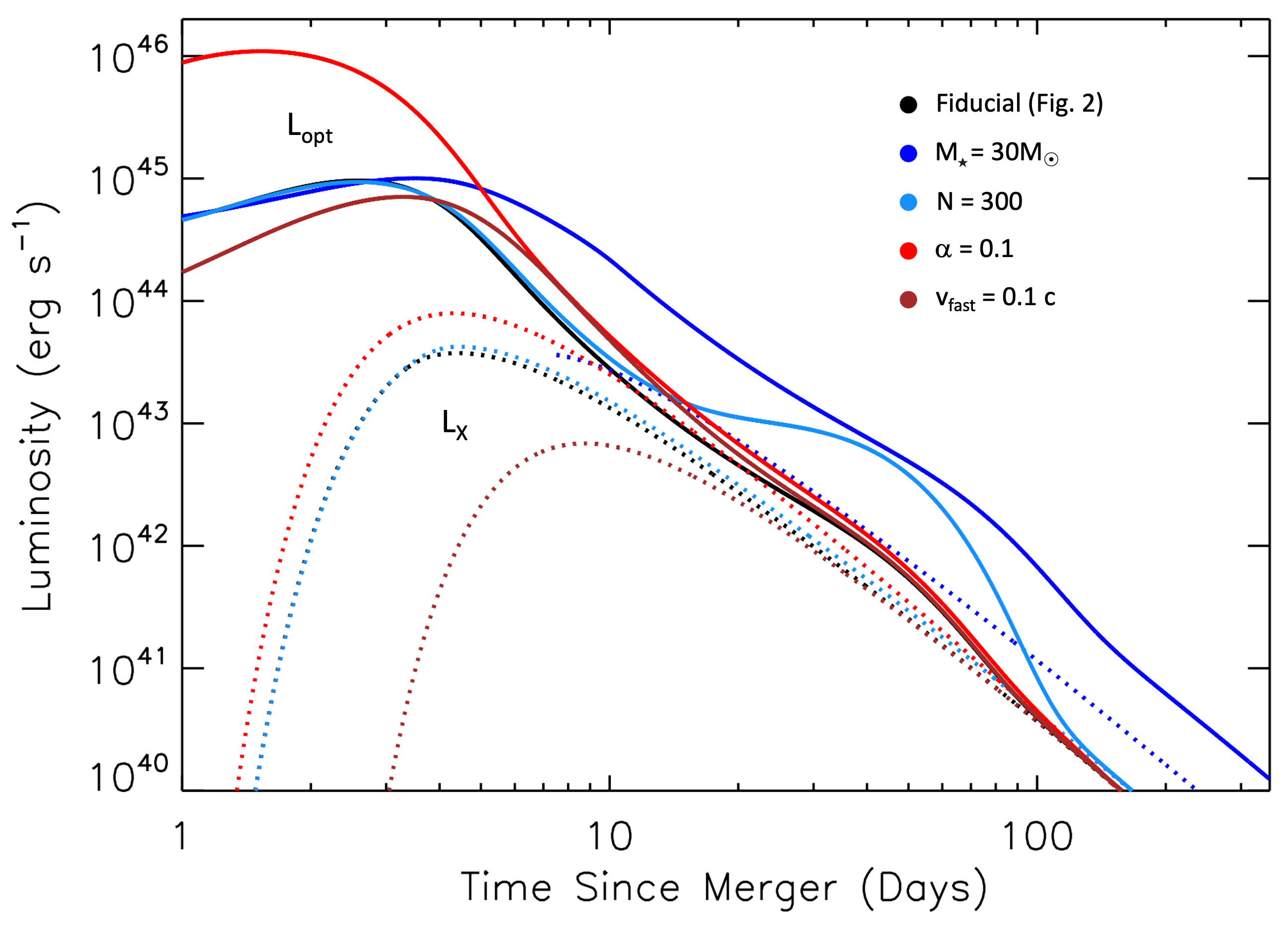}
    \caption{Similar to Figure \ref{fig:example}, but showing the effect on the total optical (solid lines) and X-ray (dotted lines) light curves of tripling the disrupted star mass $(M_{\star} = 30M_{\odot}$; dark blue), tripling the number of orbits over which pre-dynamical ejecta is released ($N = 300$; light blue), tripling the disk viscosity ($\alpha = 0.1$; red), and halving the speed of the fast ejecta component ($v_{\rm fast} = 0.1$ c; brown) relative to the fiducial model ($M_{\star} = 10M_{\odot}; \alpha = 0.03; N = 300, v_{\rm fast} = 0.1$ c; black line, Fig.~\ref{fig:example}).}
    \label{fig:examples}
\end{figure}

Figure \ref{fig:examples} shows the effect on the total optical and X-ray light curves of changing different parameters of the model relative to their values in the fiducial model (which roughly fits AT2018cow; Fig.~\ref{fig:example}).  Increasing the mass of the disrupted star increases the duration of the transient.  Increasing the duration of the pre-dynamical mass-loss phase prior to the merger ($N$ orbits) makes the shock-powered component of the optical light curve more pronounced.  Increasing the effective viscosity of the disk $\alpha$ speeds up the accretion evolution, resulting in a greater peak luminosity.  Decreasing the velocity of the fast polar ejecta delays the optical peak slightly but has a more pronounced effect on the rise-time and peak luminosity of the X-ray emission. 

\subsection{Shock-Powered Synchrotron Radio Emission}    
\label{sec:radio}

As established by previous works (e.g., \citealt{Ho+19,Margutti+19,Ho+20,Coppejans+20,Nayana&Chandra21,Bright+21,Ho+21}), the bright synchrotron radio and millimeter emission from LFBOTs is generated by shock interaction between the fast polar ejecta component $v_{\rm fast} \sim 0.1-0.5$ c and dense CSM \citep{Margalit&Quataert21}.  In particular, to explain the luminosity and duration of the radio light curves in several LFBOTs requires a particle density as high as $n \sim 10^{5}$ cm$^{-3}$ on scales of $\sim 3\times 10^{16}$ cm.  

Only in extreme cases will the spreading circumbinary disk left over from the CE or original RLOF mass-transfer phase extend to such radii $\gtrsim 10^{16}$ cm (Eq.~\ref{eq:RCD}), while fine-tuning of the merger time would be required to place the unbound ejecta from the first mass-transfer/CE phase on a similar radial scale in multiple LFBOT events (Sec.~\ref{sec:CE}); furthermore, both such CSM sources are expected to be concentrated in the binary equatorial plane and hence would not be directly impacted by the fast polar disk-wind ejecta.  

As discussed in Section \ref{sec:CSM}, outflows from the remnant circumbinary disk due to photoevaporation by the WR star, naturally generate a wind of density $n_{\rm ph} \sim 10^{5}$ cm$^{-3}$ (Eq.~\ref{eq:nph}) on the characteristic radial scale $\sim few \times R_{\rm g} \sim 3\times 10^{16}$ cm (Eq.~\ref{eq:Rg}).  The irradiated circumbinary disk and its outflows will furthermore extend well out of the disk midplane, even becoming quasi-spherical on large scales $\gtrsim R_{\rm g}$ \citep{Hollenbach+94}, such that the fast polar ejecta from the merger could interact with this material.  

The slow, photoevaporation driven wind $\dot{M} \sim \dot{M}_{\rm ph} \sim 10^{-4}M_{\odot}$ yr$^{-1}$ will interact strongly with the faster $\gtrsim 1000$ km s$^{-1}$ radiation-driven wind from the central WR star (of mass-loss rate $\dot{M} \sim 10^{-5}M_{\odot}$ yr$^{-1}$; \citealt{Nugis&Lamers00}), leading to mass entrainment and significant acceleration of the former on radial scales $\sim R_{\rm g} \sim 10^{16}$ cm.  This wind-wind interaction could rise to the inferred steepening of the radial density profile in some LFBOTs from $n \propto r^{-2}$ to $n \propto r^{-3}$ around this radial scale (e.g.~\citealt{Ho+21,Bright+21}).

\section{Connecting LFBOTs to Type Ibn/Icn SNe and Other Related Transients}
\label{sec:unification}

\begin{figure*}
    \centering
    \includegraphics[width=1.0\textwidth]{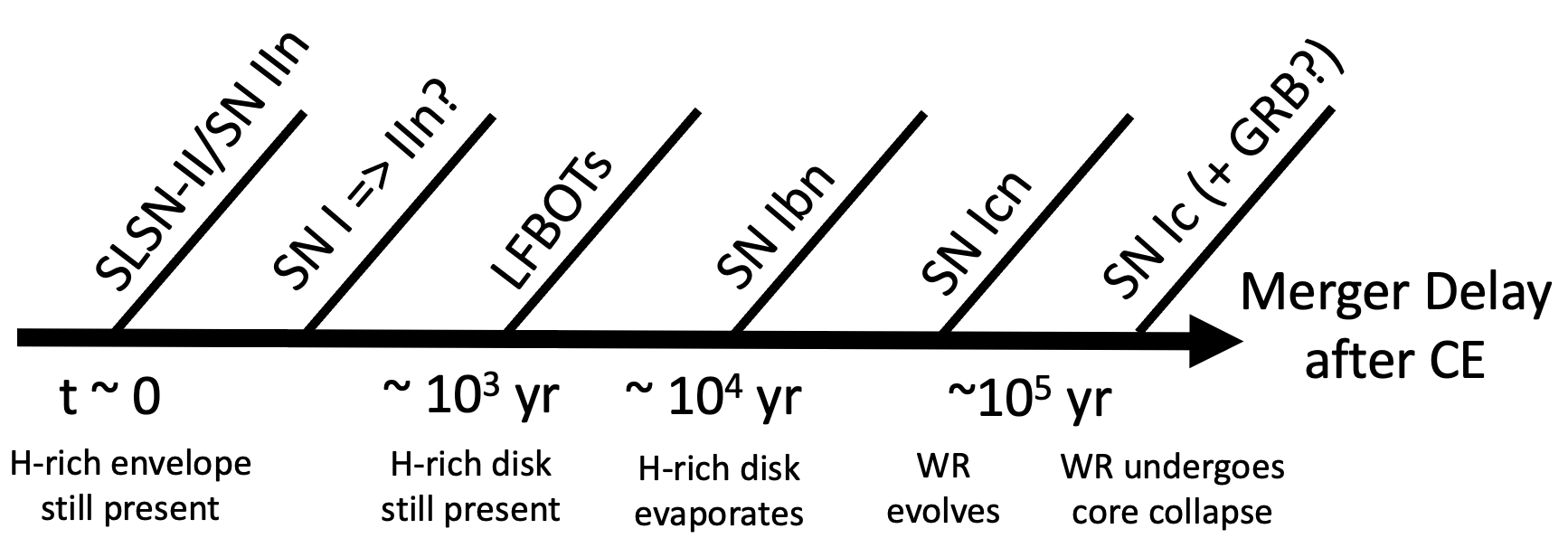}
    \caption{Unification scheme for different SN-like transients generated by the merger of a WR-BH/NS binary, as a function of the time delay between the merger and the first RLOF/CE event responsible for originally tightening the binary.  Mergers with very short delays (e.g., $0 \lesssim t \lesssim 100$ yr) will give rise to explosions still embedded in the H-rich CE envelope which may appear as a Type II SLSN or Type IIn SN (e.g., \citealt{Chevalier12}) or a Type I SN that transforms into Type IIn.  Mergers with intermediate delays (e.g., $10^{2}$ yr $\lesssim t \lesssim 10^{4}$ yr) may give rise to LFBOTs surrounded by H-depleted CSM due to the presence of a bound relic disk and its photoevaporation driven outflow (Sec.~\ref{sec:CSM}).  For mergers with longer delays (e.g., $10^{4}$ yr $ \lesssim t \lesssim 10^{5}$ yr) the relic disk has completely accreted or evaporated and the transient will appear as a Type Ibn/Icn, with the CSM interaction powered by interaction with the earliest phase of the merger ejecta.  For the longest delays $t \gtrsim 10^{5}$ yr the WR star will have evolved to a core-collapse prior to the merger, giving rise to a stripped envelope Type Ic SN, which depending on the binary separation may power a GRB-like high energy transient (e.g., \citealt{Rueda&Ruffini12,Fryer+14}).  The merger delay time may in turn depend on the mass and radial extent of the cirumbinary disk left over from the CE event (e.g., Eq.~\ref{eq:tvisc} and surrounding discussion).}
    \label{fig:timeline}
\end{figure*}

Our discussion and modeling has been focused on LFBOTs such as WR-BH/NS mergers with a large delay $\sim 10^{2}-10^{4}$ yr following the original mass-transfer or CE phase.  However, there exist striking similarities between the optical properties from LFBOTs (e.g., extremely luminous; too fast rising and decaying to be powered by $^{56}$Ni decay; evidence for H-depleted CSM interaction) and the rare class of luminous stripped envelope stellar explosions which also exhibit evidence for strong CSM interaction$-$Type Ibn (e.g., \citealt{Foley+07,Immler+08,Pastorello+15,Hosseinzadeh+17,Pellegrino+22,Maeda&Moriya22}) and Type Icn SNe \citep{Gal-Yam+22,Perley+22}.

In SNe Ibn, the CSM velocities inferred from the widths of the He lines are comparable to those of WR star winds \citep{Foley+07,Pastorello+08}, naturally leading to a preferred progenitor scenario involving exploding WR stars which exhibit abnormally high mass-loss rates just prior to core-collapse.  However, similar expansion velocities $\sim v_{\rm esc}$ (Eq.~\ref{eq:vL2}) and CSM mass and radial extent $\sim 10^{14}$ cm (Eq.~\ref{eq:RL2}) accompany the runaway mass-loss phase predicted to occur leading up to a WR-BH/NS merger (Sec.~\ref{sec:CSM}).  There are also both observational hints (e.g., \citealt{Sanders+13,Pastorello+15,Hosseinzadeh+17}) and theoretical suggestions (e.g., \citealt{Woosley17,Renzo+20,Leung+20}) that Type Ibn could arise from multiple progenitor channels.  

Type Icn SNe are a new and extremely rare class of stellar explosions which exhibit strong, narrow lines with profiles similar to seen in SNe Ibn, but originating from C, O, and other alpha elements rather than He \citep{Gal-Yam+22,Perley+22}.  \citet{Perley+22} estimate the rate of Type Icn SNe to be $\sim 0.005-0.05\%$ of the CCSNe rate, broadly overlapping (albeit with large uncertainties) the rate of LFBOTs.  One member of this class, SN 2021csp, exhibited an optical light curve similar to AT2018cow (\citealt{Perley+22,Fraser+21}; see Fig.~\ref{fig:example}) and also spectroscopic evidence for strong interaction between fast ejecta $\sim 0.1 c$ and slower $v \sim 2000-4500$ km s$^{-1}$ H/He-depleted CSM of limited radial extent $\lesssim 400R_{\odot}$ surrounding the explosion.   

Deep optical limits on SN 2021csp at late times point to a low $^{56}$Ni mass of $\lesssim 10^{-2}M_{\odot}$ \citep{Perley+22}, unless the total ejecta mass is very low.  Such a low $^{56}$Ni mass is difficult to explain in the context of a successful core-collapse explosion given the large kinetic energy of the explosion (e.g., \citealt{Woosley+02}).  A ``failed'' initial explosion which nevertheless produces a BH surrounded by an accretion disk of sufficient size (radius $\gtrsim R_{\odot}$) to generate a disk-wind explosion of long enough duration to power an LFBOT (enabled in our scenario by the binary merger), would require the progenitor WR star possess an angular momentum at collapse in tension with current stellar evolution predictions (e.g., \citealt{Fuller+19}) and far greater than even required to explain long-duration gamma-ray bursts (e.g., \citealt{MacFadyen&Woosley99}).

We propose an alternative scenario.  Motivated by the inference that the ejecta properties of SNe Ibn and Icn mirror those of distinct WR spectroscopic subtypes (He/N-rich WN versus He-poor, C-rich WC stars, respectively; \citealt{Gal-Yam+22}), we suggest that the merger-driven destruction$-$rather than core-collapse explosion$-$of WR stars with different envelope structure give rise to a Type Ibn/Type Icn dichotomy.  Conversely, mergers with shorter post-CE delay times than FBOTs will give rise to explosions more completely embedded in the hydrogen envelop from the original RLOF/CE phase and will appear spectroscopically as luminous Type II or Type IIn SNe \citep{Chevalier12} or perhaps energetic Type Ic SNe that later transform into Type IIn (e.g., \citealt{Chugai&Chevalier06,Chen+18}).  This speculative unification scheme is outlined in Figure \ref{fig:timeline}.

AT2018cow exhibited stronger evidence for hydrogen in its spectra than Type Ibn/Icn SNe, albeit at a depleted level compared to Type II or IIn SNe.  This would arise naturally if both LFBOTs and (some) Type Ibn/Icn are triggered by similar merger-driven explosions, but which emerge into different CSM environments due to a range of merger delay times after the original mass-transfer phase or CE event.  The H-rich relic circumbinary disk from this early phase provides a CSM source surrounding the explosion, but it is eventually dispersed by accretion and photoevaporative mass-loss on a timescale $\lesssim 10^{4}$ yr  (Sec.~\ref{sec:binarydiskoutflows}) and hence H would be less prevalent surrounding mergers with sufficiently long delays.

Another observational distinction between LFBOTS and Type Ibn/Icn SNe is that the radio and X-ray luminosities of the latter class are significantly lower, with X-ray/radio upper limits on SNe 2021csp for example at least an order of magnitude below those detected from AT2018cow \citep{Perley+22}.  The luminous radio/mm emission from LFBOTs arises from shock interaction on large radial scales $\lesssim 10^{15}-10^{16}$ cm with CSM which in our scenario results from the outflows of a relic circumbinary disk (Sec.~\ref{sec:radio}).  Because of the disk dispersal process, the lack of bright radio/mm emission from events with long merger (Type Ibn/Icn), would arise naturally.  

The optical emission from CSM interaction and engine reprocessing should be relatively isotropic.  However, the X-ray emission$-$which must escape through the low-density polar region of the ejecta$-$will likely be geometrically beamed.  Its luminosity may therefore depend sensitively on viewing angle, with polar observers observing brighter X-ray emission compared to those viewing the binary off-axis.  A prediction of the unification scenario would therefore be the eventual discovery of luminous X-ray emission from a subset of Type Ibn/Icn supernovae.

\section{Conclusions}
\label{sec:conclusions}

We have developed a model for LFBOTs from the binary merger of a WR star with a BH or NS companion in the delayed aftermath of a massive star CE event.  Our conclusions can be summarized as follows.
\begin{itemize}
    \item Although many progenitor models for LFBOTs have been proposed in the literature, most are challenged to simultaneously explain all of their properties (Table \ref{tab:models}), particularly the presence of a highly energetic central compact object (NS or BH) and highly aspherical $^{56}$Ni-poor ejecta interacting with massive but H-poor CSM covering a large range of radial scales $\sim 10^{14}-10^{16}$ cm surrounding the explosion. What is often lacking is a compelling causal connection between these individually atypical properties.

\item Among the few surviving models is the tidal disruption and merger of a WR star with a BH or NS binary companion, which occurs after some long delay $\gtrsim 100-1000$ years after the original stable mass-transfer or CE phase responsible for birthing the binary.  This differs from previously proposed CE core merger-driven LFBOT models (e.g., \citealt{Soker+19,Schroder+20}) which envision a prompt post-CE merger and$-$at least for the massive stars of interest$-$are more likely to produce a Type II SN from immediate interaction with the massive hydrogen envelope of the donor star \citep{Chevalier12}.

\item The tidal disruption of the WR star by a BH/NS creates an accretion disk surrounding the latter with a viscous accretion timescale of $\lesssim$ days (Eq.~\ref{eq:tvisc2}), commensurate with the peak durations of LFBOT light curves.  Such hyper-accreting flows naturally generate outflows with a wide range of velocities as a result of different launching radii in the disk, from $v_{\rm slow} \sim 3000$ km s$^{-1}$ for the bulk of the mass $\sim M_{\star} \sim 10M_{\odot}$ which emerges at low latitudes from the outer regions of the disk, up to $v_{\rm fast} \gtrsim 0.1$ c for the fastest polar outflows with lower mass $\sim 0.1M_{\odot}$ from the inner edge of the disk, both consistent with observations of LFBOTs including AT2018cow.  

The accretion rate which reaches the central compact object exhibits a normalization and decay-rate $\dot{M}_{\bullet} \propto t^{-\alpha}$ with $\alpha \approx 2$ (Eqs.~\ref{eq:Lacc}, \ref{eq:Lacc2}), consistent with the engine power required to explain AT2018cow \citep{Margutti+19} for an assumed radiation/jet efficiency $\eta \sim 10^{-2}$ similar to those predicted by GRMHD simulations of super-Eddington accretion.  Depending on how efficiently disk outflows extract angular momentum from the disk, a gradual exponential steeping in the accretion rate may occur at late times (Eq.~\ref{eq:Lacc2}), potentially consistent with the observed steepening of the X-ray light curves of AT2018cow and AT2020xnd at around 1 month.  

The bulk of the ejecta will be the unprocessed material comprising the disrupted WR star.  However, moderate quantities of intermediate mass elements like $^{20}$Ne, $^{24}$Mg, $^{28}$Si, and $^{40}$Ca may be synthesized in the hot accretion disk midplane and carried into the disk-wind outflows.  Only the very innermost radii of the disk reach sufficiently high temperatures $\gtrsim 4\times 10^{9}$ K to synthesize Fe-peak elements (Eq.~\ref{eq:T0}), so the expected yield of $^{56}$Ni in the disk outflows is predicted to small $\lesssim 10^{-2}M_{\odot}$ (Sec.~\ref{sec:burning}), consistent with the optical light curves of FBOTs and some Type Ibn/Icn SNe.  

\item We present a toy model for the optical and X-ray light curves of FBOTs, the results of which are compared in a fiducial case to data for AT2018cow (Fig.~\ref{fig:example}).  

The early-time optical light curves in WR-BH/NS merger-driven transients are mainly powered by reprocessing of X-ray emission from the inner accretion flow or jet by the fast polar ejecta, consistent with earlier inferences \citep{Margutti+19}.  As the polar ejecta becomes transparent and an increasingly large fraction of the X-rays escape, the relative X-ray luminosity contribution compared to the optical increases.  The X-ray variability timescale also naturally shortens because of weaker photon diffusion-time induced filtering once $\tau_{\rm es} \lesssim 1$.  

The accretion rate onto the BH/NS will reach the Eddington rate on a timescale of years or less.  The approach to sub-Eddington accretion could be accompanied by softening of the X-ray spectra in analogy to X-ray binary state transitions and flattening of the decaying X-ray light curve (as the radiative efficiency increases to match that of a \citealt{Novikov&Thorne73} thin-disk).  Optical/UV emission may also be generated from the outer accretion flow and lasting for years (Eqs.~\ref{eq:Rtr},\ref{eq:Ttr}).  

\item Shock interaction between the bulk of the slower wind ejecta and equatorially-concentrated CSM also plays a role in powering the observed optical emission, particularly on timescales of a few weeks after the merger, consistent with the Type Ibn-like spectral features of LFBOTs (e.g., \citealt{Fox&Smith19}).  Motivated by observations and modeling luminous red novae from ordinary stellar mergers (e.g., \citealt{Pejcha+17}), one likely source for the H-depleted CSM surrounding the merger-driven explosion are outflows from the WR star which possess typical velocities $\sim 10^{3}$ km s$^{-1}$ similar to the escape speed of the binary and occur starting tens or hundreds of binary orbital periods prior to its final tidal disruption.  

\item A promising mechanism to instigate a post-CE binary merger with a sufficiently long delay (yet still within the lifetime of the WR star), is gradual angular momentum extraction by a circumbinary disk left over from the first mass-transfer/CE phase (Sec.~\ref{sec:delayed}).  The timescale of the merger delay timescale is sensitive to the properties of the leftover CE disk, with more radially-extended and less massive disks producing longer delays (Eq.~\ref{eq:tvisc}).  The same relic disk and its outflows (driven by photoionization from the WR star) generate CSM out to a critical radial scale $\sim 10^{16}$ cm (Eq.~\ref{eq:Rg}) with the required density (Eq.~\ref{eq:nph}) to explain the bright radio/mm synchrotron emission from FBOTs via shock interaction of the fast disk-wind ejecta (Sec.~\ref{sec:radio}).

\item WR-BH/NS mergers which occur with different delays following the first mass-transfer/CE phase will be characterized by drastically different CSM environments and hence may manifest with a diverse range of observational properties (Fig.~\ref{fig:timeline}).  We speculate that some Type Ibn/Icn SNe are intrinsically ``merger-driven explosions'' similar to LFBOTs (with the Ibn or Icn class, depending on the evolutionary state of the WR being disrupted), but with merger delay times which exceed the lifetime of the circumbinary disk, resulting in little or no H-rich CSM.  If early stages of the merger process strip material from the WR star tens or hundreds of orbits before the final tidal disruption, this H-poor CSM source would be present also in Type Ibn/Icn, consistent with the strong shock interaction signatures of the latter.  This unification scheme predicts the future discovery of Type Ibn/Icn with LFBOT-like X-ray emission from the central engine, for events viewed close to the rotation axis.  

\item The massive star binaries that give rise to WR-BH/NS mergers and luminous transients like LFBOTs are in some sense nature's ``failed'' attempt to create a tight compact object binary capable of becoming a powerful gravitational wave source.  The local rate of LFBOTs and Type Icn SN (e.g., \citealt{Coppejans+20,Ho+21b}), are comparable (within the large uncertainties) to the local rate of neutron star and black hole binary mergers detected by LIGO/Virgo \citep{Abbott+21b,Abbott+21c}.  A better understanding of these transient events and their connections could therefore offer unique insights into gravitational wave source populations.    
    
\end{itemize}

\acknowledgements

I am grateful to Raffaella Margutti for providing data on AT2018cow.  I also thank Ryan Chornock, Zoltan Haiman, Jakub Klencki,  Tatsuya Matsumoto, Raffaella Margutti, Ondrej Pejcha, and particularly Mathieu Renzo for helpful discussions and suggestions on a first draft of the text.  I thank the anonymous reviewer for helpful comments that improved the manuscript. This work is supported in part by NASA (grant 80NSSC20K1557) and the National Science Foundation (grants AST-2009255, AST-2002577).


\end{document}